\documentclass[fleqn,usenatbib,useAMS]{mnras}
 \usepackage{hyperref}
\usepackage{graphicx}	
\usepackage{amsmath}	
\usepackage{amssymb}	
\usepackage{stfloats}
\usepackage{bm}
\usepackage{ulem}
\usepackage{ae,aecompl}
\usepackage{enumerate}

\newcommand{\zoh}{12+\log \mathrm{(O/H)}}

\newcommand{\irx}{L_{\rm IR}/L_{\rm UV}}

\defcitealias{Qin2019a}{Q19}
\title[Understanding the universal IRX relation]{Understanding the Universal Dust Attenuation Scaling Relation of Star-Forming Galaxies}
\author[J. Qin et al.]{
Jianbo~Qin,$^{1}$
Xian~Zhong~Zheng,$^{1,2}$\thanks{xzzheng@pmo.ac.cn}
Stijn~Wuyts,$^{3}$
Zongfei~Lyu,$^{1,2}$
Man~Qiao,$^{1,2}$
Jia-Sheng~Huang,$^{4,5}$
\and
Feng~Shan~Liu,$^{5}$
Antonios~Katsianis, $^{6}$
Valentino~Gonzalez,$^{7}$
Fuyan~Bian,$^{8}$
Haiguang~Xu,$^{9}$
\and
Zhizheng~Pan,$^{1}$
Wenhao~Liu,$^{1}$
Qing-Hua~Tan,$^{1}$
Fang~Xia~An,$^{1}$
Dong~Dong~Shi,$^{1}$
Yuheng~Zhang,$^{1,2}$
\and
Run~Wen,$^{1,2}$
Shuang~Liu$^{1,2}$
and Chao~Yang$^{1,2}$
\\ 
$^{1}$Purple Mountain Observatory, Chinese Academy of Sciences, 10 Yuanhua Road, Nanjing 210023, China\\
$^{2}$School of Astronomy and Space Science, University of Science and Technology of China, Hefei 230026, China\\
$^{3}$ Department of Physics, University of Bath, Claverton Down, Bath BA2 7AY, UK \\
$^{4}$Chinese Academy of Sciences South America Center for Astronomy, National Astronomical Observatories, \\ \ \,Chinese Academy of Sciences, Beijing 100101, China \\
$^{5}$CAS Key Laboratory of Optical Astronomy, National Astronomical Observatories, Chinese Academy of Sciences, Beijing 100101, China \\ 
$^{6}$School of Physics and Astronomy, Sun Yat-sen University, Zhuhai Campus, 2 Daxue Road, Xiangzhou District, Zhuhai 519082, China \\
$^{7}$Departamento de Astronom\'ia, Universidad de Chile, Camino del Observatorio 1515, Las Condes, Santiago 7591245, Chile \\
$^{8}$European South Observatory, Alonso de Cordova 3107, Casilla 19001, Vitacura, Santiago 19, Chile \\
$^{9}$School of Physics and Astronomy, Shanghai Key Laboratory for Particle Physics and Cosmology, \\  \,
 Shanghai Jiao Tong University, Shanghai 200240, China}

\date{Accepted 2023 December 26. Received 2023 December 20; in original form 2023 August 21}
\pubyear{2023}
\begin{document}
\label{firstpage}
\pagerange{\pageref{firstpage}--\pageref{lastpage}}
\maketitle

\begin{abstract}
	Star-forming galaxies (SFGs) adhere to a surprisingly tight scaling relation of dust attenuation parameterized by the infrared excess (IRX$\equiv$$\irx$), being jointly determined by the star formation rate (SFR), galaxy size ($R_{\rm e}$), metallicity ($Z$/Z$_\odot$) and axial ratio ($b/a$). We examine how these galaxy parameters determine the effective dust attenuation and give rise to the universal IRX relation, utilizing a simple two-component star-dust geometry model in which dust in the dense and diffuse interstellar medium (ISM) follows exponential mass density profiles, connected with but not necessarily identical to the stellar mass profiles. Meanwhile, empirical relations are adopted to link galaxy properties, including the gas--star formation relation, the dust-to-stellar size relation, as well as the dust-to-gas ratio versus metallicity relation.
By fitting a large sample of local SFGs with the model, we obtain the best-fitting model parameters as a function of metallicity, showing that the two-component geometry model is able to successfully reproduce the dependence of IRX on SFR, $R_{\rm e}$, $b/a$ at given $Z$/Z$_\odot$, as well as the dependence of power-law indices on metallicity. 
Moreover, we also retrieve constraints on the model geometry parameters, including the optical depth of birth clouds (BCs), BC-to-total dust mass fraction, BC covering factor of UV-emitting stars, and star-to-total dust disc radius ratio, which all evolve with galaxy metallicity. Finally, a consistent picture of how the star-dust geometry in SFGs evolves with galaxy metallicity is discussed.   
\end{abstract}


\section{Introduction} \label{sec1}

Dust, as a crucial component of the interstellar medium (ISM),  plays a significant role in driving the cycles of baryons and energy.  Firstly,  it acts as a channel for converting radiation pressure into mechanical energy and outflows \citep{Thompson2015, Naddaf2022}. Secondly, dust serves as an important coolant for star and planet formation, and an interface for the formation of molecules  \citep{Galliano2018}.  Additionally, dust has a profound influence on the spectral energy distributions (SED) of galaxies by absorbing the stellar ultraviolet (UV) to optical radiation and  thermally re-emitting at the infrared (IR) wavelengths. This alteration of the SED significantly affects galaxy observables \citep{Calzetti2000, Conroy2013}. Dust attenuation, defined as the effective sight-line absorption of light by dust in a galaxy (also  referred to as `obscuration'),\footnote{In contrast, `extinction' refers to the absorption and scattering of the light from a point source by dust along the line of sight.}  depends on both the dust content and the geometry between dust and stars. It is closely intertwined with star formation, chemical enrichment and structural growth of the galaxy.  Understanding dust attenuation and the regulation by dust of physical processes across cosmic time is crucial not only for improving measurements of intrinsic galaxy properties, but also for unravelling the connections between dust, gas, metals, and stars in accordance with the structural evolution of galaxies \citep{Salim2020}.

For star-forming galaxies (SFGs), various fundamental scaling relations involving gas, metals, and stars have been established up to a redshift of $z\sim 10$, revealing their rapid evolution across this redshift range. One of these scaling relations is the star formation main sequence, which shows a nearly linear increase in star formation rate (SFR) with stellar mass ($M_\ast$) at a given redshift, while globally decreasing in SFR as redshift decreases \citep{Noeske2007, Wuyts2011, Whitaker2012, Guo2013, Speagle2014, Schreiber2015, Katsianis2020, Thorne2021, Leja2022, Kouroumpatzakis2023, Popesso2023}. The density of star formation in SFGs is regulated by the density of cold gas, following the Kennicutt-Schmidt (K-S) Law ($\Sigma_{\rm gas}$--$\Sigma_{\rm SFR}$) \citep{Kennicutt1998}. 
Another important scaling relation is the stellar mass--metallicity ($M_{\ast}$--$Z$) relation, which reveals an increase in gas-phase metallicity with stellar mass among SFGs. However, this relation also shows a rapid decline in metallicity
 at fixed $M_{\ast}$ as redshift increases 
\citep{Tremonti2004, Erb2006, Mannucci2010, Zahid2012, Zahid2014, Maiolino2019, Sanders2021, Nakajima2023, Curti2023}. In addition, SFGs are predominantly characterized by disc morphologies, and their sizes, quantified by the half-light radius ($R_{\rm e}$), typically increase with stellar mass. The stellar mass--size ($M_{\ast}$--$R_{\rm e}$) relation of SFGs demonstrates a rapid size increase over time for SFGs with a given $M_{\ast}$ \citep{van der Wel2014a, Suess2019, Mowla2019, Nedkova2021, Ono2023, van der Wel2023}.  These scaling relations provide a framework for understanding the evolution of and interplay between gas, metals, stars, and structural properties of SFGs from high redshifts to the present day.

On the other hand, dust attenuation serves as an indicator of effective dust columns in SFGs, which is closely related to gas density and metallicity. Observationally, dust attenuation in SFGs is often quantified using the IR to UV luminosity ratio, known as the IR excess (IRX$\equiv$$\irx$). Generally, SFGs with higher star formation rates (SFR) tend to exhibit higher infrared luminosities and dust temperatures \citep{Dale2002, Rieke2009, Schreiber2018}, resulting in increased dust attenuation \citep{Whitaker2014}. The dust attenuation in SFGs has been found to positively correlate with stellar mass, SFR, surface density of SFR, UV continuum slope, and gas-phase metallicity \citep{Kong2004, Martin2005, Johnson2007, Garn2010, Zahid2013, Battisti2016}. However, investigations on local SFGs have revealed that the dependence of dust attenuation on SFR (as well as galaxy inclination angle) weakens with decreasing metallicity \citep{Wild2011, Xiao2012}. Similarly, \citet{Zahid2017}  found distinct scaling relations of dust attenuation with SFR at different stellar masses. In the high-mass regime ($M_\ast>10^{10.2}$\,M$_\odot$), dust attenuation increases with SFR, while in the low-mass regime  ($M_\ast<10^{10.2}$\,M$_\odot$), it decreases with SFR. \citet{Whitaker2014} found that the relation between galaxy stellar mass and attenuation evolves moderately at $M_\ast>10^{10.5}$\,M$_\odot$ over $0<z<2.5$,  but displays no or little evolution in the low-mass regime over the  examined redshift range.   Several studies using different indicators of attenuation have reported this puzzling `no evolution' phenomenon in the stellar mass--dust attenuation relation for SFGs \citep{Price2014, Whitaker2017, McLure2018, Shapley2022, Shapley2023, Zhang2023}.  These correlations between dust attenuation and galaxy properties in SFGs demonstrate that dust attenuation is a complex process with multiple factors at play.

In the study by \citet[hereafter referred to as \citetalias{Qin2019a}]{Qin2019a}, we utilized a large sample of local SFGs to investigate the correlation between IRX and other galaxy parameters. By separating these parameters and determining independent correlations, we obtained several key findings:  (1) Our analysis contradicted the previous understanding that galaxy stellar mass is the primary factor influencing IRX. We discovered that dust obscuration is not correlated with galaxy stellar mass, once controlling for other, more fundamental drivers.
(2) We found that dust attenuation, as indicated by IRX, is determined by a combination of various parameters including SFR (or the indicator of infrared luminosity $L_{\rm IR}$), half-light radius ($R_{\rm e}$), gas-phase metallicity ($Z$), and galaxy inclination (axial ratio $b/a$). The relationship between these parameters and IRX can be described by a power-law function:  $IRX=10^\alpha\,L_{\rm IR}^{\beta}\,R_{\rm e}^{-\gamma}\,(b/a)^{-\delta}$ with the power-law slopes  $\beta$, $\delta$, and $\gamma$ decreasing (i.e. getting closer to zero) as the metallicity decreases.

Furthermore, we verified that the empirical relation of dust attenuation obtained from local SFGs also holds out to $z\sim 2$, demonstrating its universality. This universal IRX relation provides insights into the evolution of galaxies in terms of star formation, chemical enrichment, and morphological structure. However, the mechanism underlying how these three physical processes interact to make galaxies follow this relation remains to be explored. The determination of IRX is influenced by the dust content, star formation activity, and the geometry of dust and stars within a galaxy. It is now evident that the geometry of dust and stars is crucial for understanding the differences between various attenuation measurements and for quantifying the relative contributions that determine the overall dust attenuation of a galaxy.

In the past three decades, much effort has been devoted to studying galaxy dust attenuation using star-dust geometry models \citep{Calzetti1994}. Two commonly considered geometries are a uniform foreground dust screen and a homogeneous mixture of dust and emitting sources. However, in most cases, these simple geometries fail to adequately describe the observed attenuation patterns in real galaxies. \citet{Li2019} investigated the spatially resolved Balmer decrement distribution and its relation to other local and global properties, and concluded that neither the foreground screen nor the homogeneous mixture geometry can accurately represent the observed attenuation patterns \citep[see also][]{Kreckel2013}. In addition, \citetalias{Qin2019a}  pointed out that while the simple homogeneous geometry can reproduce the IRX relations at Solar metallicity, it fails to reproduce the flat slopes observed at lower metallicities. This suggests that there is no universal star-dust geometry that applies to all types of galaxies. Instead, a more complex and flexible star-dust geometry model is required to better match the observations.

One approach to achieving this is by allowing the relative sizes of the dust and stellar discs to freely vary in the geometry model \citep{Xiao2012}. If the stars are more centrally concentrated than the dust disc, they will suffer an enhanced attenuation since the optical depth of the dust disc decreases with galactocentric radius.  Another way to address the complexity of star-dust geometry is by considering a `two-component' dust geometry, where the ISM of galaxies consists of dense birth clouds (BCs; associated with short-lived stars) and the surrounding diffuse ISM \citep{Charlot2000}. This two-component geometry successfully explains the difference in attenuation between nebulae (birth clouds) and old stars in galaxies \citep{Charlot2000, Wild2011, Price2014, Koyama2019, Shivaei2020}. The BCs in a galaxy follow certain distributions of ages and initial masses. If these distributions are similar for each galaxy, it would be expected that the average attenuation of BCs does not depend on either SFR or inclination. On the other hand, IRX measures the attenuation of the total UV light from young and intermediate-age populations. The former reside in the BCs while the latter are no longer surrounded by birth clouds due to stellar feedback. Then the steepness of IRX correlations might be influenced by the relative importance of dust attenuation caused by BC and diffuse dust.  

In this work, we aim to build a new geometry model based on the two-component prescription with flexible model parameters, to reproduce the observational trends of IRX correlations highlighted in \citetalias{Qin2019a}. We focus on the scaling relations of dust attenuation, in particular, on the systematic trends of the steepness of IRX scaling relations with galaxy metallicity. In Section~\ref{sec2}, we outline the details of our model as well as the impact of model parameters on dust attenuation. In Section~\ref{sec3}, we describe how to fit the observed data with our geometry model. We show best-fitting results in Section~\ref{sec4} and discuss the implications in Section~\ref{sec5}. Finally, we summarize our main results in Section~\ref{sec6}.

\section{The star-dust geometry model} \label{sec2}

\subsection{The distribution of dust and stars, and dust opacity} \label{sec2.1}

While star formation does occur in various types of galaxies, including irregular galaxies and elliptical galaxies, the highest rates of ongoing star formation are typically observed in the most abundant disc galaxies which provide better conditions for the birth of stars. For example, ellipticals were found to contribute only 10-13\% of the total ongoing star-formation budget \citep{Zheng2009, Wuyts2011, Nelson2013, van der Wel2014b, Kaviraj2014, Lofthouse2017}. We thus focus on disc SFGs and address how dust attenuation is affected by different quantities/processes. We aim to examine the scaling relations of dust attenuation among galaxy populations. It is reasonable to assume that stars and dust are well mixed in the ISM, and the size distribution of dust grains and their chemical compositions are identical everywhere across a galaxy. 

The ISM in a galaxy can be briefly described with two components: the dense birth clouds (BCs, or star-forming regions) with short-lived young stars, as well as more extended large-scale distributions of diffuse ISM, dust and old stars \citep{Charlot2000,Wild2011}. The large-scale distribution of diffuse dust plays a major role in mediating the propagation of photons in galaxy discs and, at least in nearby galaxies, dominates the bolometric output of dust emission. To simplify, the total dust optical depth can be considered as the sum of the optical depth caused by the \textbf{diffuse dust (diff)} plus the \textbf{dense BC dust (dense)}:
\begin{align}
	\tau=\tau^{\rm diff}+\tau^{\rm dense}.   
\end{align}
This calculation relies on the assumption that the BCs and young stars share the same geometry distribution and no overlapping effects between BCs along the line of sight are taken into account in our model.

A double exponential form is commonly used in astrophysics to capture the observed distribution of matter in galactic discs \citep{Misiriotis2000, Smith2015}.
For the diffuse components, the distribution of diffuse dust $\rho_{\rm d}$ and stars $\rho_\star$, both residing in a disc configuration, can be described by a double exponential, respectively  
\begin{align}
	\rho_{\rm d}(r,h)=&\rho_{\rm d}(0,0)\exp\left({-\frac{r}{R_{\rm d}}}{-\frac{|h|}{H_{\rm d}}}\right) ,\\ 
	\rho_\star(r,h)=&\rho_\star(0,0)\exp\left({-\frac{r}{R_\star}}{-\frac{|h|}{H_\star}}\right) ,
\end{align}
where $r$ and $h$ are the radius and height in cylindrical coordinates, while $R$ and $H$ are the scale-length and scale-height of the dust or stellar disc, respectively. We use subscript $\star$ to represent the stellar disc and use subscript $d$ to represent the diffuse dust disc hereafter.  We note that the subscript $d$ always refers to the diffuse dust disc. To not be confused, we add superscript $tot$ if denoting the total dust disc (including also BCs). Then the integrated diffuse dust mass is given as 
\begin{align} \label{int_dust_mass}
	M_{\rm d}&=\int_0^\infty \int_{-\infty}^{\infty} \rho_{\rm d}(r,h) 2\pi r dr dz \nonumber \\	
		&=\int_0^\infty 2 \pi  r dr \int_{-\infty}^{\infty}\rho_{\rm d}(0,0)\exp\left(-\frac{r}{R_{\rm d}}-\frac{|h|}{H_{\rm d}}\right) dh \nonumber \\
		&=4\pi\rho_{\rm d}(0,0) R_{\rm d}^2 H_{\rm d} .
\end{align}
  
Similarly, the integrated intrinsic luminosity of stellar emission is 
\begin{align} \label{eq_lint}
	L_{\rm \star,int}&=\int_0^\infty \int_{-\infty}^{\infty} \rho_\star(r,h) 2\pi r dr dh \nonumber \\	
		     &=\int_0^\infty 2 \pi r dr\int_{-\infty}^{\infty}\rho_\star(0,0)\exp\left(-\frac{r}{R_\star}-\frac{|h|}{H_\star}\right) dh \nonumber \\
		     &=4\pi\rho_\star(0,0) R_\star^2 H_\star . 
\end{align}

The observed luminosity of stellar emission due to the dust attenuation is 
\begin{align} \label{eq_lobs}
	L_{\rm \star,obs}=&\int_0^\infty\,2 \pi r\,dr\,\int_{-\infty}^{\infty}\rho_\star(0,0)\exp\left[-\frac{r}{R_\star}-\frac{|h|}{H_\star}-\tau_{\rm rh}\right] dh ,
\end{align}
where $\tau_{rh}$ is the diffuse optical depth along the line of sight towards a given ($r$, $z$). For a galaxy observed under a face-on orientation,     
\begin{align} \label{eq_tau_rh}
	\tau_{\rm rh}&=\int_h^\infty \kappa \rho_{\rm d}(0,0)\exp\left(-\frac{r}{R_{\rm d}}-\frac{|h|}{H_{\rm d}}\right) dh \nonumber \\
		&=\frac{\kappa \Sigma_{\rm d}}{4}\times\begin{cases}
                    	    	e^{-r/R_{\rm d}}(2-e^{h/H_{\rm d}}), & \text{if $h<0$} ,\\
			    	e^{-r/R_{\rm d}}e^{-h/H_{\rm d}}, & \text{if $h>0$} , 
		                \end{cases}
\end{align}
where $\Sigma_{\rm d}$ is the diffuse dust surface density defined as $M_{\rm d}/(\pi R_{\rm d}^2)$, and $\kappa$ is the dust mass extinction coefficient that converts the dust surface/column density into dust optical depth. $\kappa$ is wavelength dependent because different wavelengths of light have different absorption and scattering cross sections for the same dust grain, e.g. UV light has higher $\kappa$ than optical. Here we omit the subscript $\lambda$ to avoid the complexity of the notation hereafter. 

By definition, the effective optical depth caused by the diffuse dust is 
\begin{equation}\label{eq_tau_diff}
\begin{split}
\tau^{\rm diff}&= -\ln\left(\frac{L_{\star,\rm obs}}{L_{\star,\rm int}}\right)  \\
		   & = -\ln\int_0^\infty \frac{r dr}{2R_\star ^2}\int_{-\infty}^{\infty}\exp\left[-\frac{r}{R_\star}-\frac{|h|}{H_\star}-\tau_{rh}\right] \frac{dh}{H_\star} , \\ 
\mathrm{where} &  \\ 
	&\tau_{\rm rh}=\frac{\kappa\Sigma_{\rm d}}{4}\begin{cases}
                    	    	e^{-r/R_{\rm d}}(2-e^{h/H_{\rm d}}), & \text{if $h<0$} , \\
			    	e^{-r/R_{\rm d}}e^{-h/H_{\rm d}}, & \text{if $h>0$} .
			\end{cases} 
\end{split}
\end{equation}
We then do the following variable substitutions: 
\begin{equation}\label{eq_vs}
\begin{split}
 & r^\prime=r/R_\star , \\ 
 & h^\prime=h/H_\star ,\\  
 & \hat{R}=R_\star/R_{\rm d}, \\  
 & \hat{H}=H_\star/H_{\rm d}. 
\end{split}
\end{equation}
Equation~\ref{eq_tau_diff} can be rewritten as   
\begin{equation}\label{eq_tau_rh_prime}
\begin{split}
\tau^{\rm diff} & =-\ln\int_0^\infty \frac{1}{2}r^\prime dr^\prime\int_{-\infty}^{\infty}\exp\left[-r^\prime-|h^\prime|-\tau^\prime_{rh}\right] dh^\prime ,  \\  
\mathrm{where}  &   \\ 
 & \tau^\prime_{\rm rh}=\frac{\kappa\Sigma_{\rm d}}{4} \begin{cases}
		e^{-\hat{R} r^\prime}(2-e^{\hat{H} h^\prime}), & \text{if $h^\prime<0$} ,\\ 
		e^{-\hat{R} r^\prime}(e^{-\hat{H} h^\prime}), & \text{if $h^\prime>0$} .
			\end{cases} 
\end{split}
\end{equation}

The diffuse dust attenuation $\tau^{\rm diff}$ in our geometry model is only determined by the diffuse dust surface density $\Sigma_{\rm d}$, star-to-diffuse dust scale-length ratio $\hat{R}$, star-to-diffuse dust scale-height ratio $\hat{H}$, and the dust mass extinction coefficient $\kappa$. We notice that the thickness of both the dust and stellar discs (i.e., $H_{\rm d}/R_{\rm d}$ and $H_\star/R_\star$) have vanished.

The above equations allow translating a given diffuse dust mass and geometry to the effective optical depth seen by a face-on observer.  Under an inclined angle, sightlines to individual disc regions will cross dust located at different radii, but as we will demonstrate in Appendix~\ref{sec_c}, the associated net attenuation of the galaxy light can still be approximated adequately using Equations~\ref{eq_tau_diff}-\ref{eq_tau_rh_prime}, simply by boosting $\Sigma_d$ by a factor $1/(b/a)$.

For the dense (or BC) component, we first assume that the BCs are spherical and have similar sizes. We return to this assumption in Section~\ref{sec5.1}. Additionally, for two similar size BCs, more metal-rich BCs with higher dust-to-gas ratio are expected to have higher BC optical depth ($\tau_{\rm bc}$). The galaxy dust attenuation probed by IRX measures the attenuation of the total UV light from young and intermediate-age populations (up to a few $\times10^8$\,yr). The young stars are surrounded by BCs while the intermediate-age stars no longer live in BCs (due to feedback) but still contribute the UV emission \citep{Kennicutt2012}. Given the evolutionary picture of BC dispersal, it is natural to introduce a BC covering factor $C_{\rm bc}$, defined as the total fraction of UV emission that is subject to BC attenuation. We note that our covering factor $C_{\rm bc}$ is not the same as the clumpiness factor in the literature \citep[e.g.][]{Tuffs2004,vanderGiessen2022}, they assume the UV light is totally obscured in BCs (i.e., $\tau_{\rm bc}\gg1$) whereas we do not. The optical depth caused by the dense BCs follows 
\begin{align}\label{tau_dense}
	\tau^{\rm dense}&=-\ln\left[\frac{I_{\rm \star,int}(1-C_{\rm bc})+C_{\rm bc}\times I_{\rm \star,int} \exp(-\tau_{\rm bc})}{I_{\rm \star,int}}\right] \nonumber \\
			    &=-\ln\left[1-C_{\rm bc}+C_{\rm bc}\times \exp(-\tau_{\rm bc})\right] , 
\end{align}
where $I_{\rm \star,int}$ denotes the UV light intrinsically emitted by the stellar population.

Another important parameter in our two-component model is the BC dust mass fraction $F_{\rm bc}$, defined as the BC-to-total dust mass fraction. It controls the proportion of each component in the two-component model, and hence the evolution of dust geometry. For example, at $F_{\rm bc}=0$ and $F_{\rm bc}=1$, it reverts back to the pure BC-dominated and  diffuse dust-dominated geometry, respectively. Naturally, $F_{\rm bc}$ also impacts the relative contribution of dust attenuation caused by BC and diffuse dust. Using the BC mass fraction $F_{\rm bc}$, we are able to link the diffuse dust surface density $\Sigma_{\rm d}$ in Equation~\ref{eq_tau_rh_prime} with the observationally more relevant total dust surface density $\Sigma_{\rm d}^{\rm tot}$. The latter is defined as $\Sigma_{\rm d}^{\rm tot}\equiv M_{\rm d}^{\rm tot}/[\pi (R_{\rm d}^{\rm tot})^2]$, where $M_{\rm d}$ is the total dust mass and $R_{\rm d}^{\rm tot}$ is the scale-length of the total dust disc. Then we have 
\begin{align}
	\Sigma_{\rm d}=(1-F_{\rm bc})\frac{M_{\rm d}^{\rm tot}}{\pi R_{\rm d}^2}&=(1-F_{\rm bc})\frac{M_{\rm d}^{\rm tot}}{\pi (R_{\rm d}^{\rm tot})^2}\times \left(\frac{R_{\rm d}^{\rm tot}}{R_{\rm d}}\right)^2\nonumber \\ 
								       &=(1-F_{\rm bc})\Sigma_{\rm dust}^{\rm tot}\times \left(\hat{R}^{\rm tot}\right)^2 , 
\end{align}
where $\hat{R}^{\rm tot}= R_{\rm d}^{\rm tot}/R_{\rm d}$. In Appendix~\ref{sec_a}, we explain how the size ratio of the total to diffuse dust disc can be written as a function of $\hat{R}$ and $F_{\rm bc}$. 
Then the total dust optical depth can be obtained  with
\begin{align}\label{eq_tau_tot}
	\tau^{\rm tot}=\tau^{\rm diff}\left(\kappa\Sigma_{\rm d}^{\rm tot},F_{\rm bc}, \hat{R}, \hat{H}\right)+\tau^{\rm dense}\left(\tau_{\rm bc}, C_{\rm bc}\right) . 
\end{align}

In summary, the effective dust optical depth in our two-component model depends on the dust extinction coefficient $\kappa_\lambda$, total dust surface density $\Sigma_{\rm _{\rm d}}^{\rm tot}$, BC-to-total dust mass fraction $F_{\rm bc}$, star-to-diffuse dust disc scale length ratio $\hat{R}$ and scale height ratio $\hat{H}$, BC optical depth $\tau_{\rm bc}$, and BC covering fraction $C_{\rm bc}$. Other parameters have vanished during the derivation of the above formula. 
 
Although the derivation of the above equations is based on young/intermediate-age stellar discs, it also holds for stellar discs at different ages (or a nebular disc) with appropriate changes to the model parameters. The old stellar disc has longer wavelength emission than young stars and therefore has smaller dust opacity (smaller $\kappa$ and $\tau_{\rm bc}$). On the other hand, the old stars are not expected to be surrounded by BCs, and their BC coverage fraction $C_{\rm bc}$ should be set to 0. Conversely, the nebular emission traces $<10^7$yr stellar emission, thus younger than those traced by the UV (or IRX), and is expected to have a higher $C_{\rm bc}$.  

While in this paper, we focus on the IRX diagnostic of attenuation, the framework outlined in this section will thus also be applicable to the interpretation of, for example, Balmer decrement measurements, and how they contrast to IRX \citep[see e.g.][]{Qin2019b}.

\begin{figure*}
\centering
\includegraphics[width=\textwidth]{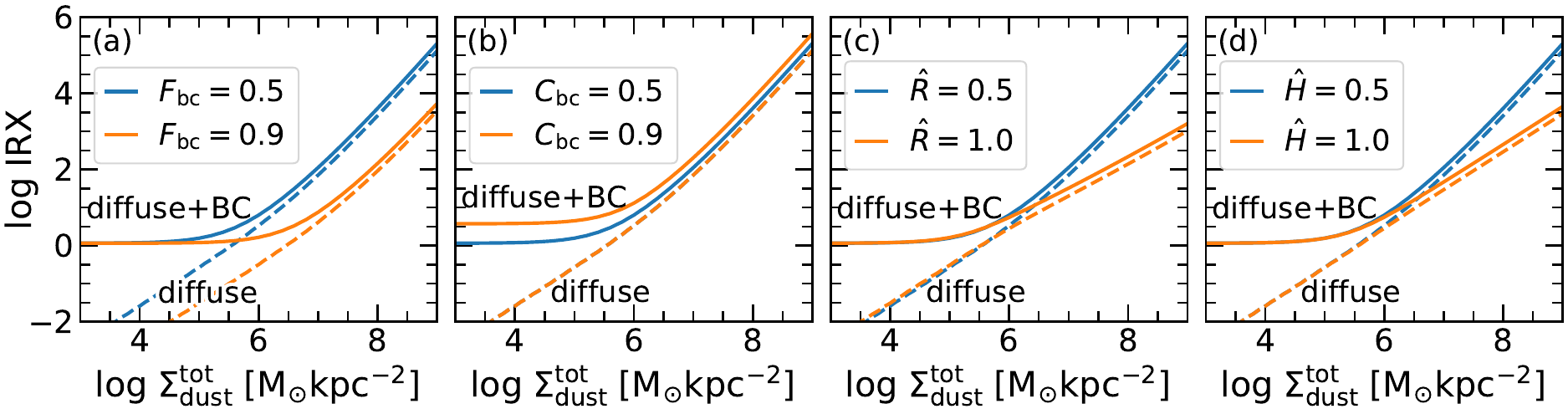}
\caption{IRX as a function of $\Sigma_{\rm dust}^{\rm tot}$ at different $F_{\rm bc}$ (a), $C_{\rm bc}$ (b), $\hat{R}$ (c), and $\hat{H}$ (d), indicated by blue and orange colours. The dashed lines represent the dust attenuation caused by diffuse dust, while the solid lines consider both diffuse and BC dust attenuation (with a fixed $\tau_{\rm bc,V}=0.5$). All model parameters ($F_{\rm bc}$, $C_{\rm bc}$, $\hat{R}$, $\hat{H}$) are fixed to a typical value of 0.5 except for the one indicated in the diagram. }
\label{fig:fig1}
\end{figure*}

\subsection{Dependence of IRX on model parameters}\label{sec2.2}

The equations derived in Section~\ref{sec2.1} give the optical depth $\tau^{\rm tot}$ of a galaxy, while the attenuation indicator of infrared excess (IRX$\equiv\irx$), defined as the ratio between infrared and UV luminosity, is used in our observational analysis. The intrinsic UV emission is dominated by the young ($<10$\,Myr) and intermediate-age (10--500\,Myr) stellar populations. For a constant star formation history (SFH), the young stellar populations  contribute  56\% of the total UV luminosity (1216\AA--3000\AA), 19\% from these at ages of $10-30$\,Myr, and the remaining 25\% stems mainly from stellar populations of ages 30--500\,Myr.  Note that dust heating by old stars can contribute an increasing fraction to the far-IR luminosity (i.e., cold dust emission) at decreasing specific SFR but the fraction is marginal for normal SFGs \citep{Nersesian2019}.  Moreover,  the total IR luminosity of local SFGs used in our analysis is derived from the mid-IR luminosity  (see Section~\ref{sec3.1}) in which the contribution from  the dust heated by old stars is negligible. We thus assign the total IR luminosity to the dust-reprocessed radiation of young and intermediate-age stellar populations, and take IRX to reflect their dust attenuation.

A conversion between IRX and FUV optical depth should be made. Following \citet{Hao2011}, the IRX can be obtained by  $IRX=\left(e^{\tau_{\rm FUV}}-1\right)/0.46$, where $\tau_{\rm FUV}$ is the dust optical depth at the FUV according to Equation~\ref{eq_tau_tot}.  This conversion is based on the energy balance principle in the sense that the absorbed UV-to-optical radiation is totally transformed into the IR through dust thermal emission. Throughout our analysis, we adopt a fixed $\kappa_{\rm V}$ of $0.62\times 10^{-5}$\,M$_\odot^{-1}$\,kpc$^{2}$ following \citet{Kreckel2013} (see Section~\ref{sec3.4}).  We further adopt the \citet{Calzetti2000} attenuation curve, which has $\tau_{\rm FUV}/\tau_V=2.5$. 
Here the dust optical depths are associated with the projected dust surface density and thus dependent on viewing angle.  Since our IRX estimates are based on the computed $\tau_{\rm FUV}$, they too will tightly correlate with the projected dust column density $\Sigma_{\rm dust,proj}$ (see also Appendix~\ref{sec_c}).   We caution that in detail, the application of the energy balance principle is only warranted when comparing the UV-to-optical emission versus IR emission as integrated over 4$\pi$ steradian.  For individual viewing angles, deviations from energy balance may occur as the IR radiation tends to be emitted isotropically, whereas the emerging UV light is not.  The latter effect is not captured in our modelling.  For disc SFGs,  the average attenuation over 4$\pi$ steradian equals that of an inclined SFG with project axial ratio $b/a=0.6$ (see \citetalias{Qin2019a}).  Accounting for the fact that the IR emission is anticipated to emerge isotropically would modestly increase (decrease) IRX for galaxies with higher (lower) $b/a$ than 0.6 compared to the estimates obtained with the approach adopted in this work.
Variations in the effective attenuation law, that in practice may arise from star-dust geometry and viewing angle effects, are also not captured.  For instance, for the fixed $\kappa_{\rm V}$ adopted in this paper, a greyer (i.e., flatter) attenuation law would give rise to a lower $\tau_{\rm FUV}$ and thus reduced IRX.  We return to this point in Section~\ref{sec_5.4}. 

Fig.~\ref{fig:fig1} shows the model-predicted IRX as a function of $\Sigma_{\rm dust}$ for a set of model parameters. If $\tau_{\rm bc}=0$, i.e., in the case of diffuse dust only, IRX increases monotonically with dust surface density. This is consistent with the prediction of a homogeneous mixture star-dust geometry given by \citetalias{Qin2019a}, although the quantitative form of the relation shows modest differences due to the non-uniform density distribution of dust and stars (see also figure~1 in \citealt{Zhang2023}). Once BC attenuation is considered (where for illustrative purposes we here adopt a $V$-band BC optical depth of $\tau_{\rm bc,V} = 0.5$), the $\Sigma_{\rm dust}$--$IRX$ relation flattens at the low-$\Sigma_{\rm dust}$ end. This is because the BC attenuation is assumed to be constant and independent of the opacity of diffuse dust. From Panel (b) we can see that the attenuation caused by cloud dust depends not only on the adopted BC opacity but also on the BC covering fraction $C_{\rm bc}$. The latter controls the fraction of UV emission that arises from within BCs. The more UV stars are located in BCs, the more dust attenuation. 

The attenuation caused by diffuse dust is determined by $F_{\rm bc}$, $\hat{R}$, and $\hat{H}$ but independent of $C_{\rm bc}$ at a given total dust surface density $\Sigma_{\rm dust}$. $F_{\rm bc}$ controls the remaining fraction of diffuse dust. The higher $F_{\rm bc}$, the less diffuse dust and hence the lower the overall dust attenuation. $\hat{R}$ and $\hat{H}$ control the relative spatial distribution between UV-emitting stars and dust in the galaxy. With decreasing $\hat{R}$ and $\hat{H}$, the UV emission becomes more centrally concentrated within the dust disc, both radially and vertically. For an exponential dust disc, the central region always has a higher dust column density. We note that $\hat{R}$ and $\hat{H}$ have a very similar impact on the resulting $\Sigma_{\rm dust}$--IRX relation, and consequently might not be very distinguishable in subsequent model fitting. We will impose $\hat{R}=\hat{H}$ in our model as discussed in Section~\ref{sec3.4}.  

In conclusion, in the BC dominated regime, IRX increases with $\tau_{\rm bc}$ and $C_{\rm bc}$, and has a flat $\Sigma_{\rm dust}$--IRX relation; in the diffuse dust dominated regime, IRX decreases with $F_{\rm bc}$, $\hat{R}$ and $\hat{H}$, and has a steep $\Sigma_{\rm dust}$--IRX relation. The slope of the $\Sigma_{\rm dust}$--IRX relation is thus not constant over the full range of dust surface densities, but is controlled by the relative importance between BC and diffuse dust attenuation. We notice that the (projected) dust surface density is expected to correlate with the metallicity, SFR, $R_{\rm e}$ and $b/a$ \citep{Wuyts2011,Li2019,Shapley2022}. This gives a hint that the systematic trends between the power-law slope of IRX as a function of $L_{\rm IR}$ or SFR, $R_{\rm e}$ and $b/a$ with gas-phase metallicity as shown in \citetalias{Qin2019a} may arise from systematic changes in the relative importance of BC attenuation.  

\section{Fitting the local SFGs with the geometry model}\label{sec3} 

\subsection{The local SFGs and universal IRX relation } \label{sec3.1}

We carry out our investigation of dust attenuation using the sample and data from \citetalias{Qin2019a}, which consists of 32,354 local SFGs. More details about the sample selection and data extraction can be found in \citetalias{Qin2019a}. We note that in \citetalias{Qin2019a} the IR luminosities (8--1000\,$\mu$m) were estimated from the single {\it WISE} W4 band. However, we find that the estimated total $L_{\rm IR}$ from the combination of W3 and W4 bands have a smaller dispersion than using a single W4 band (see Appendix~\ref{sec_b}). In this work, we therefore update the total IR luminosity by combining the {\it WISE} W3 and W4 bands. The IR-to-UV luminosity ratio is referred to as IRX and SFR is also estimated from the combination of IR and UV luminosities following \citet{Bell2005}.

The original universal IRX relation states that IRX is determined by a combination of the infrared luminosity $L_{\rm IR}$,  half-light radius ($R_{\rm e}$), gas-phase metallicity ($Z$) and galaxy inclination (axial ratio $b/a$). In this work, we replace the observational parameter of IR luminosity by the physically more relevant parameter of star formation rate (SFR). The two parameters are tightly correlated with each other (see figure~2 of \citetalias{Qin2019a}).  The best-fitting relation of IRX as a function of $Z$, SFR, $R_{\rm e}$ and $b/a$ obeys
\begin{equation}\label{eq:eq4}
\begin{split}
IRX & =10^\alpha\, \left(\frac{SFR}{\rm M_\odot \rm yr^{-1}}\right)^{\beta}\,\left(\frac{R_{\rm e}}{\rm kpc}\right)^{-\gamma}\,(b/a)^{-\delta} ,  \\
\mathrm{where} &   \\ 
& \alpha=1.46\log(Z/\rm Z_\odot)+0.92 ,  \\
& \beta=0.67\log(Z/\rm Z_\odot)+0.55 ,  \\
& \gamma=0.95\log(Z/\rm Z_\odot)+0.80 ,  \\
& \delta=1.52\log(Z/\rm Z_\odot)+1.00 .  
\end{split}
\end{equation}
Here, $\log (Z/$Z$_\odot)=\zoh-8.69$ represents the oxygen abundance and $\rm Z_\odot$ refers to Solar metallicity. The power-law slopes decrease with decreasing metallicity. Moreover, galaxies at $z\sim 2$ also follow this empirical relation for dust obscuration, supporting its universality.

\subsection{Determining projected $\Sigma_{\rm dust}$ from metallicity, SFR, $R_{\rm e}$ and b/a} \label{sec3.2}

We have shown that IRX in the model is controlled by the dust surface density and other model parameters. However, our empirical universal IRX relation shows that IRX depends on metallicity, $L_{\rm IR}$ or SFR, size and axial ratio. We infer dust surface density from $Z$, SFR, $R_{\rm e}$, and $b/a$ in the following steps. At first, dust surface density ($\Sigma_{\rm dust}=M_{\rm dust}/(\pi R_{\rm e}^2)$) can be inferred from gas surface density via the dust-to-gas ratio (DGR) as 
\begin{align} \label{eq_sigdust}
	\Sigma_{\rm dust}=\Sigma_{\rm gas}\times DGR .
\end{align} 
The gas surface density is found to be tightly correlated with SFR surface density, known as the Kennicutt-Schmidt Law
\begin{align} \label{eq_ks}
	\Sigma_{\rm SFR}= \phi \Sigma_{\rm gas}^n ,   
\end{align}
where $n\sim$1.4 as suggested by \citet{Kennicutt1998}. Here we do not use a fixed value for K-S slope but let it vary freely in our model. We notice that the `surface density' presented here is defined using a half-light radius,\footnote{Here we do not distinguish strictly between the half-light radius and half-mass radius.  The SDSS sample only provides optical measurements of the half-light radii, tracing the stellar emission albeit subject to potential light-weighting effects \citep[see e.g.][]{Zhang2023}.} while the Equations in Section~\ref{sec2.1} use disc scale-length instead. The two radii relate as $R_{\rm e}=1.678 R_\ast$ for an exponential disc, and when applying Equation~\ref{eq_tau_rh_prime} we account for this difference in definition of surface density. Motivated by the power-law relation between stellar and dust radii discussed by \citet{Mosenkov2022}, we parametrize  
\begin{align} \label{eq_rere}
	R_{\rm e,dust}= \psi R_{\rm e,star}^p . 
\end{align}
On the other hand, the dust-to-gas ratio is found to be correlated with gas-phase metallicity in the form of a power-law \citep{Issa1990,Lisenfeld1998,Leroy2011,Remy-Ruyer2014,Galliano2018,De Vis2019} as  
\begin{align} \label{eq_zdg}
	DGR=\epsilon Z^q .
\end{align}
\citet{Remy-Ruyer2014} found that the Z--DGR relation follows a linear relationship, corresponding to a constant dust-to-metal ratio, at high metallicity (12+log(O/H)$\gtrsim 8$) but breaking down in the lower metallicity regime where dust growth becomes inefficient \citep[see the review by][]{Galliano2018}. Given that our sample does not have galaxies with 12+log(O/H)$<8$, a single power-law Z--DGR relation is used in the following analysis. One should be careful when extrapolating our model to the low-metallicity regime with 12+log(O/H)$<$8. Again, the power-law slope $q$ is also a free parameter in our model.   

Putting Equations~\ref{eq_sigdust}, \ref{eq_ks}, \ref{eq_rere}, and \ref{eq_zdg} together, we have 
\begin{align} \label{eq_sigdust_final}
	\Sigma_{\rm dust}&=\left(\frac{SFR}{\pi\phi \psi^2 R_{\rm e}^{2p}}\right)^{1/n}\times \epsilon Z^q \nonumber \\ 
		      &=\mu\left[\frac{SFR}{M_{\odot}\rm yr^{-1}}\left(\frac{R_{\rm e}}{\rm 3\,kpc}\right)^{-2p}\right]^{1/n}\times \left(\frac{Z}{\rm Z_\odot}\right)^q , 
\end{align}
where $\mu\equiv\epsilon/[(\pi \phi \psi^2)^{1/n}]$ can be seen as a new normalization parameter since these sub-parameters are degenerate and indistinguishable. Finally, we interpret the axial ratio as a projection effect in determining projected dust surface density, i.e., 
\begin{align} \label{eq_sigdust_proj}
	\Sigma_{\rm dust}^{\rm proj}=\mu\left[\frac{SFR}{\rm M_{\odot}\rm yr^{-1}}\left(\frac{R_{\rm e}}{\rm 3\,kpc}\right)^{-2p}\right]^\frac{1}{n}\left(\frac{Z}{\rm Z_\odot}\right)^q \div (b/a) .  
\end{align}
Here we do not distinguish between the axial ratios of stellar and dust discs, as the difference between the two is not significant. We notice that there is a possible risk to interpreting the axial ratio as a simple projection effect if the galaxy is not an ideal thin disc. We will show that it will not significantly affect our results in Appendix~\ref{sec_c}. Finally, for brevity, we omit the `proj' superscript hereafter. 

In summary, the (projected) $\Sigma_{\rm dust}$ of galaxies in our observed sample are calculated from four observed parameters: $Z$, SFR, $R_{\rm e}$ and b/a, that used to define the universal dust attenuation relation in \citetalias{Qin2019a}.  This calculation involves four free parameters: $\mu$, $p$, $n$ and $q$.

\begin{table*} 
 \centering
 \caption{The model parameters of our geometry model} \label{tab1}
\renewcommand{\arraystretch}{1.2}
 \scalebox{1}{
 \begin{tabular}{cccc}
\hline \hline
	 Parameters                              &Description                          							&Range    &Best fit\\
\hline 
	 $\kappa_{V}$                     &\texttt{Dust mass extinction coefficient at $V$-band} (M$_\odot^{-1}$\,kpc${^2}$)          	&$0.62\times 10^{-5}$  &- \\
  \hline
	 $\log\mu$                              &\texttt{Normalization in determining $\Sigma_{\rm dust}$ (M$_\odot$\,kpc$^{-2}$) from } & 5.16 &-  \\
						& \texttt{SFR, $R_{\rm e}$, $Z$, $b/a$ in logarithm} 						      	 & 		&\\
	 $n$                            &\texttt{Power-law slope of $\Sigma_{\rm SFR}\propto \Sigma_{\rm gas}^n$}                                                     &[0, 3]  	&1.83\\
	 $p$                           &\texttt{Power-law slope of $R_{\rm e,dust}\propto R_{\rm e,star}^p$}                          &[0, 3] 	&0.69\\
	 $q$                          &\texttt{Power-law slope of $DGR\propto Z^q$ (also $\tau_{\rm bc}\propto Z^q$)} &[0, 3]       	&0.98 \\
						\hline
	 $\tau_{\rm bc, Z_\odot}$               &\texttt{Optical depth of single BC (V-band) at Z$_\odot$}                  	&[0, 3]    		&0.31\\
	 $F_{\rm bc,Z_\odot}$                  &\texttt{BC-to-total mass fraction $F_{\rm bc}$ at Z$_\odot$}             			&[0, 1]   	&0.15\\
	 $\eta$                            &\texttt{Power-law slope of metallicity dependence of $F_{\rm bc}$}                    	&[-9, 9]        &$-$1.4\\
	 $C_{\rm bc,Z_\odot}$                 &\texttt{Fraction of UV light emitted from within BCs, at Z$_\odot$}                &[0, 1]        &0.84 \\
	 $\nu$                            &\texttt{Power-law slope of metallicity dependence of $C_{\rm bc}$}                      	&[-9, 9]        &$-$3.9 \\
	 $\hat{R}$                          &\texttt{Scale-length ratio of star-to-diffuse dust disc}                                     &[0.1, 1.1]     &0.46\\
	 $\hat{H}$                          &\texttt{Scale-height ratio of star-to-diffuse dust disc}                                     &$\equiv \hat{R}$   &-\\
  \hline  \hline 
\end{tabular}}
\end{table*}

\subsection{Metallicity dependence of $\mathbf{\tau_{\rm bc}}$, $\mathbf{F_{\rm bc}}$, and $\mathbf{C_{\rm bc}}$} \label{sec3.3}

Metallicity is the most important parameter in our attenuation model, as it is believed to affect not only the dust surface density but also the star-dust geometry.  At first, the BCs in this model are assumed to be identical, i.e., have the same radii and gas masses. Their dust properties are determined by the dust-to-gas ratio of the BC, which is found to be tightly correlated with gas-phase metallicity. Following Equation~\ref{eq_zdg}, the optical depth $\tau_{\rm bc}$ can be written  as  
\begin{align}
	 &\tau_{\rm bc}=\tau_{\rm bc, Z_\odot} \times (Z/\rm Z_\odot)^q .
\end{align}

On the other hand, how to set the forms of $F_{\rm bc}$ and $C_{\rm bc}$ as a function of metallicity is a non-trivial task because there are no direct observational constraints to guide us. Some hints can be obtained from the scaling relations of other observed quantities. The specific SFR (sSFR$=SFR/M_\ast$) is found to be anti-correlated with gas-phase metallicity in the form of a power law \citep[e.g.][]{Lara-Lopez2013, Calabro2017}. Generally speaking, SFR traces recent star formation in dense ISM (associated with BC) while the $M_\ast$ traces old stars in the diffuse ISM (associated with diffuse dust). Therefore, we can assume the BC-to-diffuse dust ratio follows a power-law
\begin{align}
	 &F_{\rm bc}^{\prime}=M_{\rm d}^{bc}/M_{\rm d}^{\rm diff}=\zeta \times (Z/\rm Z_\odot)^\eta ,
\end{align}
and by definition, the BC-to-total mass ratio then follows to be
\begin{align} \label{eq_fbc}
	F_{\rm bc}&=\frac{1}{1+1/F_{\rm bc}^\prime}=\frac{1}{1+\left[\zeta \times (Z/\rm Z_\odot)^\eta\right]^{-1}} .
\end{align}
At Solar metallicity, $F_{\rm bc, Z_\odot} = 1/(1+1/\zeta)$. We rewrite Equation~\ref{eq_fbc} by replacing the $\zeta$ with $F_{\rm bc}^{\rm Z_\odot}$ as  
\begin{align} \label{eq_Z_fbc}
	F_{\rm bc}=\frac{1}{1+\left(1/F_{\rm bc,Z_\odot}-1\right) \times (Z/\rm Z_\odot)^{-\eta}} .
\end{align}
On the other hand, $C_{\rm bc}$ describes the fraction of UV light emitted from within BCs (relative to the total UV emission).  The stellar populations in a galaxy that contribute UV emission can be divided into two categories: the young stars (Y) that form in dense BCs, and the intermediate-age stars (M) that are no longer surrounded by BCs (due to the stellar feedback) but still contribute to the UV radiation. We use $Y$ and $M$ to represent the UV emission from the young and intermediate-age stars, respectively. Similar to $F_{\rm bc}$, we also assume a power-law relation between $C_{\rm bc}^\prime=Y/M$ and metallicity, and by definition,  
\begin{align}
	C_{\rm bc}=\frac{Y}{Y+M}=\frac{1}{1+\left[\xi\times (Z/\rm Z_\odot)^\nu\right]^{-1}} .
\end{align}
Similarly, we have $C_{\rm bc,Z_\odot}=1/(1+1/\xi)$ at Solar metallicity and thus have  
\begin{align} \label{eq_Z_Cbc}
	C_{\rm bc}=\frac{1}{1+\left(1/C_{\rm bc,Z_\odot}-1\right) \times (Z/\rm Z_\odot)^{-\nu}} . 
\end{align}

\begin{figure*}
\centering
\includegraphics[width=0.95\textwidth]{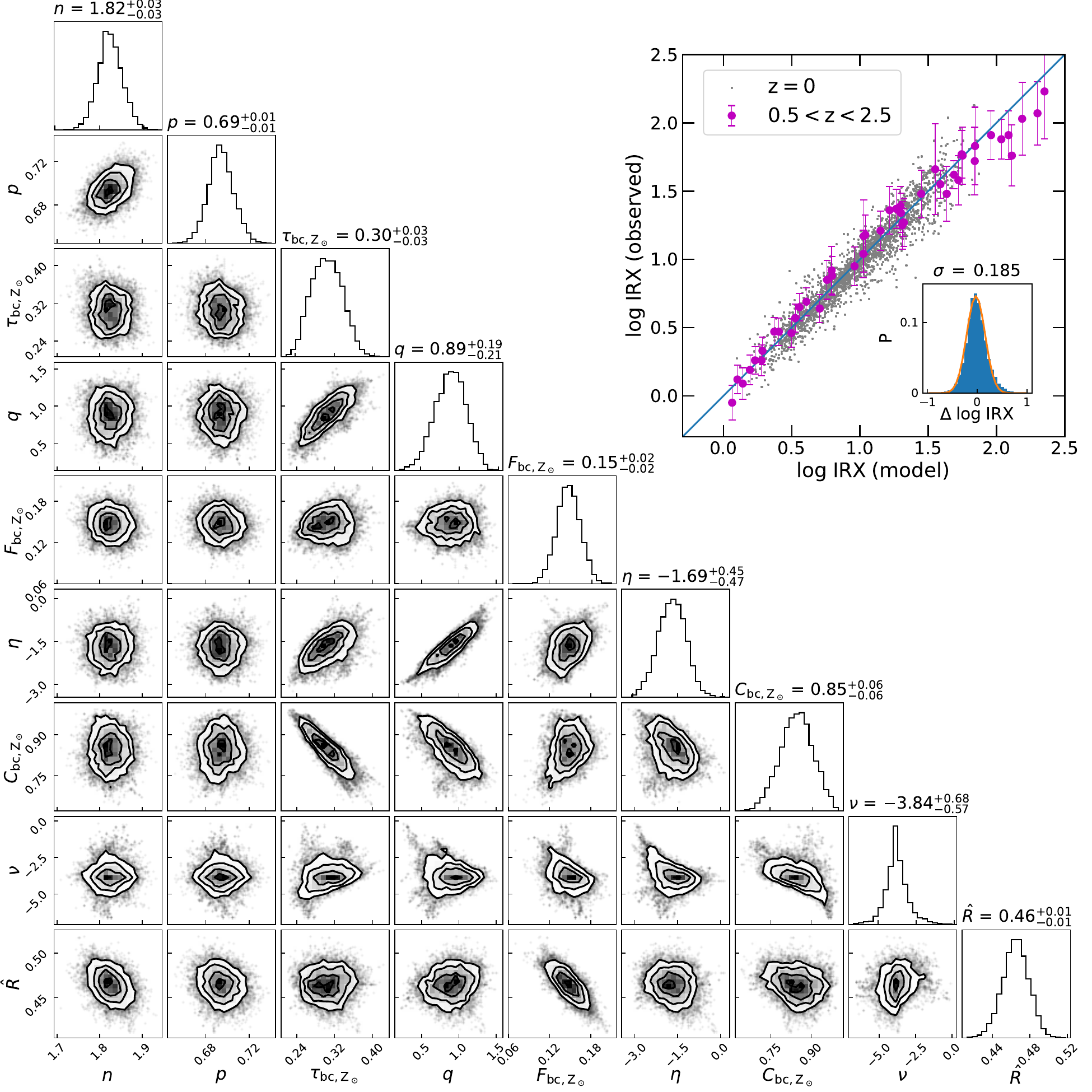}
\caption{Corner plot showing the best-fitting results of posterior PDFs of nine model parameters: $n$, $p$, $\tau_{\rm bc,Z_\odot}$ ($V$-band), $q$, $F_{\rm bc,Z_\odot}$, $\eta$, $C_{\rm bc,Z_\odot}$, $\nu$ and $\hat{R}$. The 50th percentile value (slightly different from the likelihood-weighted mean value in Table~\ref{tab1}) and 1\,$\sigma$ upper and lower uncertainties (obtained based on the 16-50-84th percentile values) are given at the top of each histogram.}
\label{fig:fig2}
\end{figure*}

\subsection{MCMC fitting and reducing the model parameters} \label{sec3.4}

In summary, our geometry model determines IRX from metallicity, SFR, $R_{\rm e}$ and $b/a$ in the following two steps: for obtaining dust surface density from the observed metallicity, SFR, $R_{\rm e}$, and $b/a$, we have the model parameters $\mu$, $n$, $p$, and $q$; for obtaining IRX from $\Sigma_{\rm dust}$, we have parameters $\kappa$, $\tau_{\rm bc,Z_\odot}$, $q$, $C_{\rm bc,Z_\odot}$, $\eta$, $F_{\rm bc,Z_\odot}$, $\nu$, $\hat{R}$, and $\hat{H}$. The definition, units and ranges of these model parameters are summarized in Table~\ref{tab1}. 

Of them, $\kappa_{\rm V}$ is fixed to $0.62\times 10^{-5}$\,M$_\odot^{-1}$\,kpc$^{2}$ according to \citet{Kreckel2013}. \citet{Kreckel2013} convert dust mass surface density to $A_{V}$ by assuming the observed Milky Way ratio of visual extinction to hydrogen column, and a fixed dust-to-gas mass ratio. They obtain $A_V/\Sigma_{\rm dust}=0.67\times 10^{-5}$\,M$_\odot^{-1}$\,kpc$^2$, corresponding to $\kappa_{V}=0.62\times 10^{-5}$\,M$_\odot^{-1}$\,kpc$^{2}$.\footnote{The small numerical difference comes from  $A_{V}=1.086\tau_{V}$.} Again, a \citet{Calzetti2000} attenuation curve is adopted with $\tau_{\rm FUV}/\tau_V=2.5$.

The normalization parameter $\mu$ in determining $\Sigma_{\rm dust}$ from metallicity, SFR, $R_{\rm e}$ and $b/a$ can be constrained in observations. The DustPedia project uses the Herschel Space Telescope to make far-infrared high-resolution observations of 810 nearby galaxies, which is an ideal sample for our $\Sigma_{\rm dust}$ calibrations from observations. By definition in Equation~\ref{eq_sigdust_final}, the normalization parameter $\mu$ is defined as the $\Sigma_{\rm dust}$ at $SFR = 1$\,M$_\odot$\,yr$^{-1}$, $R_{\rm e,star} = 3$\,kpc, and $Z=\rm Z_\odot$. We select a sub-sample with a limited range of $R_{\rm e}$ around 3\,kpc (from 2 to 4\,kpc) and 12+log(O/H) around 8.69 (from 8.6 to 8.85) and fit the sample to obtain the best-fitting power-law relation between $\Sigma_{\rm dust}$ and SFR. At $SFR=1$\,M$_\odot$\,yr$^{-1}$, we finally obtain the normalization parameter $\mu\approx 10^{5.16}$\,M$_\odot$\,kpc$^{-2}$. This value is fixed in our model fitting.

We have two structural parameters in the model: the star-to-diffuse dust scale-length ratio $\hat{R}$ and scale height ratio $\hat{H}$. By analysing the dust profile of edge-on galaxy discs, \citet{Mosenkov2022} found that both the 500-to-100\,$\mu$m scale-height ratio ($H_{500}/H_{100}$) and the scale-length ratio ($R_{500}/R_{100}$) are on average larger than unity.  Given that the 100\,$\mu$m and 500\,$\mu$m emission roughly trace SFR and dust mass, respectively, it suggests that the galaxy star-to-diffuse dust scale-height ratio $\hat{H} < 1$ and the star-to-diffuse dust scale-length ratio $\hat{R} < 1$ For simplicity, we let $\hat{H}\equiv \hat{R}$ in our model fitting.

Then, our model has the remaining nine free parameters to be constrained: $n$, $p$, $q$, $\tau_{\rm bc,Z_\odot}$, $C_{\rm bc,Z_\odot}$, $\eta$, $F_{\rm bc,Z_\odot}$, $\nu$, and $\hat{R}$. We make use of the Monte Carlo Markov chain (MCMC) sampling method to explore the 9-dimensional parameter space and efficiently characterize the uncertainties associated with the derived parameters. In detail, we consider a Gaussian likelihood function and assume flat priors for our model parameters in the following ranges: $n=[0, 3]$, $p=[0, 3]$, $q=[0,3]$, $\tau_{\rm bc,Z_{\odot}}=[0,3]$, $F_{\rm bc,Z_\odot}=[0,1]$, $\eta=[-9,9]$, $C_{\rm bc,Z_\odot}=[0,1]$, $\nu=[-9,9]$, and $\hat{R}=[0.1,1.1]$. With the large sample of our local SFGs, we can place considerable constraints on the geometry model with these free parameters. The Python package \textsc{emcee}\footnote{\url{https://emcee.readthedocs.io/en/stable/}} \citep{Foreman-Mackey2013} is used to perform the MCMC sampling using an affine invariant algorithm  proposed by \citet{Goodman2010}. The algorithm behind \textsc{emcee} has several advantages over traditional MCMC sampling methods and it has excellent performance as measured by the autocorrelation time. The sampler is initialised around the position in parameter space that gives the best fit to the data with a maximum likelihood method.

As shown in Fig.~\ref{fig:fig2}, we obtain an excellent match to the observed IRX. The 1\,$\sigma$ dispersion of the IRX residual is $\sim 0.19$\,dex, consistent with the dispersion of empirical power-law fitting applied by \citetalias{Qin2019a}. Moreover, this relation derived from local SFGs also holds for galaxies up to $z=3$. Fig.~\ref{fig:fig2} also shows the corner plot of the MCMC fit in order to examine the robustness of the best-fitting parameters.  We can see that, although there are some degeneracies (e.g. a lower $C_{\rm bc}$ tends to couple with a higher $\tau_{\rm bc}$ in the fitting and vice versa), the model parameters are well constrained. Even in the worst case (e.g. an anti-correlation between inferred $C_{\rm bc,Z_{\odot}}$ and $\tau_{\rm bc,Z_{\odot}}$), the relative error does not exceed 30\%.  We point out that our sample includes 37  local SFGs with $Z<\sim 1/3$\,Z$_\odot$ to provide reasonable constraints on these BC model parameters, although the vast majority of our sample galaxies cover the high-metallicity regime where these parameters are hardly constrained. The estimated values of each parameter are listed in Table~\ref{tab1}.

\section{Results} \label{sec4}

We show in this section the results of fitting the IRX of local SFGs with our geometry model. We first consider the best-fitting power-law indices of the ISM scaling relations in Section~\ref{sec4.1} and the best-fitting geometry parameters as a function of metallicity in Section~\ref{sec4.2}. Then, in Section~\ref{sec4.3}, we show how our geometry model reproduces the universal IRX relation presented in \citetalias{Qin2019a}.

\subsection{The best-fitting scaling relations in determining $\Sigma_{\rm dust}$ } \label{sec4.1}

We have three free parameters $n$, $p$, and $q$ in determining $\Sigma_{\rm dust}$ based on the galaxy scaling relations. The best-fitting KS-Law power-slope $n$ is 1.84, which is slightly steeper than the value of 1.4 by \citet{Kennicutt1998} or the revised version of 1.5 by \citet{Kennicutt2021}. However, the precise KS-Law slope depends on the sample selection and the method used to measure the physical quantities. For example, \citet{Kennicutt2021} found that the slope increases to 1.92 if a metallicity dependent $X_{\rm CO}$ (following the prescription of \citealt{Bolatto2013}) is adopted, consistent with our best-fitting result. Using a somewhat different but physically motivated prescription of $X_{\rm CO}$, \citet{Narayanan2012} also found a similarly steep slope of 1.95.   In a modelling of observed Balmer decrements using simplified dust geometries, sharing several ingredients in common with our approach, \cite{Li2019} likewise concluded that a significantly superlinear KS-slope (of $\sim 1.5$) was required to reproduce the observed H$\alpha$/H$\beta$ ratios.

The best-fitting relation between dust and stellar ($r$-band) radius follows $R_{\rm e,dust}\propto R_{\rm e,star}^p$ with the power slope $p=0.69$. Using the carefully selected sample from the DustPedia project, the best-fitting relation between dust and stellar radius follows $R_{\rm e,dust}\propto R_{\rm e,star}^{0.8}$, very close to our model prediction. This indicates that the growth of the dust disc is slower than the stellar disc. In an inside-out disc growth scenario, the materials that build a stellar disc are primarily accreted from the surrounding environment. In this sense, the properties of a galaxy would be to a large extent shaped by the properties of the accreted material (such as specific angular momentum, accretion rate, etc.). \citet{Pan2021} found that more gas-rich galaxies have a higher HI-to-stellar-size ratio. Our low-mass, gas-rich galaxies have a higher dust-to-stellar size ratio.    

A near linear Z--DGR relation (i.e., $q\sim1$) is suggested by our model fitting. This is consistent with the observational results that the DGR is well represented by a power law with a slope of $\sim$1 \citep[at least at 12+log(O/H)$>$8.2, e.g.][]{James2002, Draine2007a, Leroy2011, Sandstrom2013, Giannetti2017, Galliano2018}, and with theoretical expectations due to the dust grain growth \citep[e.g.][]{Asano2013,Zhukovska2014, Feldmann2015, Aoyama2017, McKinnon2018}. We notice that the linear Z--DGR relation may break below a critical metallicity \citep{Remy-Ruyer2014,Galliano2018}. \citet{Galliano2018} defined a critical metallicity regime of 12+log(O/H)=[8, 8.3]. We recall that our sample has metallicities ranging from 12+log(O/H)= 8.3 to 8.9, which is higher than the critical metallicity. On the other hand, \citet{De Vis2019} suggested that a single super-linear relation (slope $\sim$[1.78,2.45], depending on the adopted metallicity calibration) is sufficient to describe the observed Z--DGR relation at all metallicities. This is based on the chi-square ($\chi^2$) of a linear regression to the sample and does not necessarily imply that the true relationship follows this functional form. If we only focus on the high metallicity regime, as shown in figure~10 of their paper, the dust-to-metal ratio more or less flattens with metallicity, indicating that there is a linear Z--DGR relation in the high metallicity regime.

\begin{figure}
\centering
\includegraphics[width=\columnwidth]{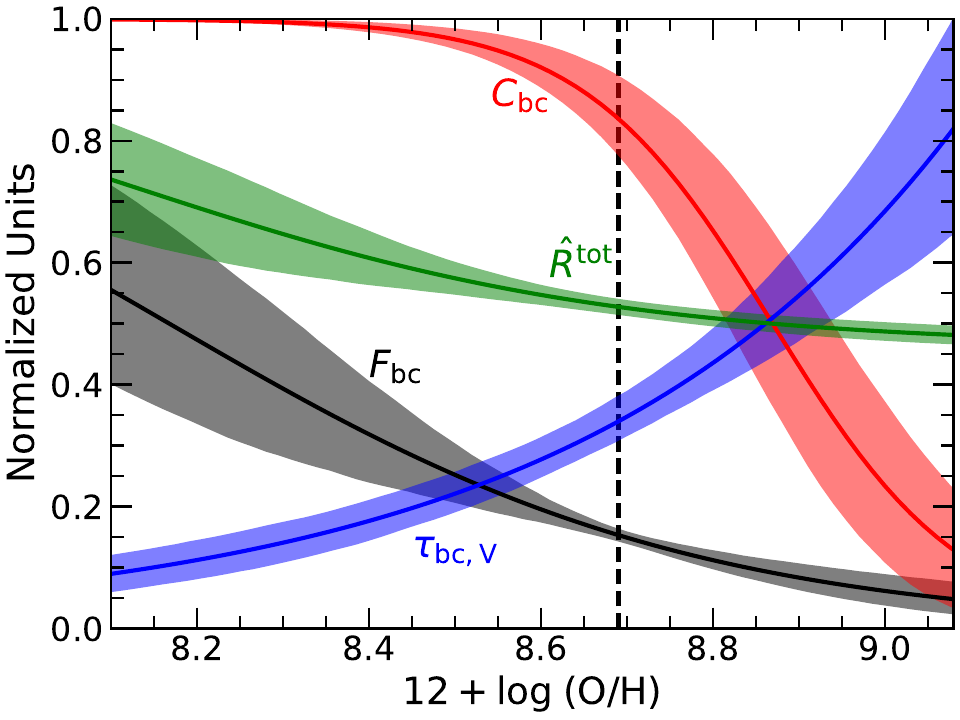}
\caption{Model parameters as a function of gas-phase metallicity.  Different colours denote the optical depth of a single BC in the $V$-band $\tau_{\rm bc,V}$ (blue), the BC-to-total mass fraction $F_{\rm bc}$ (black), the fraction of UV light emitted from within BCs $C_{\rm bc}$ (red), the star-to-total dust scale-length ratio $\hat{R}^{\rm tot}$ (green). Here $\hat{R}^{\rm tot}$ differs from the star-to-diffuse dust disc scale-length ratio $\hat{R}$ which is metallicity-independent in our model. The vertical dashed line marks Solar metallicity.}
\label{fig:fig3}
\end{figure}

\subsection{The best-fitting geometry parameters} \label{sec4.2}

For the geometry parameters, including $F_{\rm bc}$, $\tau_{\rm bc,Z_{\odot}}$, $C_{\rm bc}$, and $\hat{R}^{\rm tot}$, we show them as a function of metallicity in Fig.~\ref{fig:fig3}. We find a birth-cloud optical depth of $\tau_{\rm bc,V}=0.31$ at Solar metallicity, which is comparable to the value of 0.33 at same metallicity as given in \citet{Lu2023}. $\tau_{\rm bc,V}$ increases with metallicity following $\tau_{\rm bc}\propto Z$, i.e., a linear relation. This comes from the best-fitting linear Z--DGR relation, and the $\tau_{\rm bc}$ in our model is only determined by the dust-to-gas ratio.      

The BC mass fraction $F_{\rm bc}$ increases with decreasing metallicity. It equals 0.14 at Solar metallicity and increases to 0.53 at 1/4 Solar metallicity. This is consistent with the lower-mass/metal-poor galaxies that have younger stellar populations and more clumpy geometry \citep{Baes2020}. By modelling the inclination dependence of stellar and nebular attenuation together with a dust geometry model, \citet{Lu2022} suggested that the typical BC mass fraction is 0.3 for an MW-like galaxy. Determining $F_{\rm bc}$ in observations is a non-trivial task, as it is not a directly observed quantity. The observed dust SED can be divided into two parts: the hot dust around young stars and the diffusely distributed cold dust \citep{Draine2007a}, very similar to our definition of BC and diffuse dust, respectively. By decomposing the infrared SED of nearby galaxies, \citet{Draine2007b} found that the mass fraction of hot dust is in the range of 0-6\%. On the other hand, the molecular gas is often associated with BCs and the atomic gas is associated with the diffuse ISM. The molecular gas fraction ($\sim$ $F_{\rm bc}$) is $\sim$20\% for our local galaxies \citep{Catinella2018}. We do not intend to compare our $F_{\rm bc}$ to either hot dust fraction or molecular gas fraction, but note that they span a similar range as our best-fitting $F_{\rm bc}$.   

The BC covering fraction also decreases with metallicity, with values close to unity at lower metallicity but dropping to 0.84 at Solar metallicity. Massive and metal-rich galaxies are relatively older and have a higher fraction of intermediate-age stars that still contribute to the UV emission but had their surrounding BCs dispersed due to stellar feedback.  Similar to $F_{\rm bc}$, there are no direct observational constraints on $C_{\rm bc}$. The dense clouds  at low metallicity not only have a higher fraction of their dust comprised in BCs, but also a higher fraction of UV stars embedded in BCs. In contrast, if there is no BC dust (i.e., $F_{\rm bc}=0$), there are no UV stars surrounded by BCs (i.e., $C_{\rm bc}=0$).    
 
The radiation and stellar winds from the newly-born massive stars may clean up the leftover gas in BCs over a timescale of only a few Myr, before the occurrence of supernova (SN) explosions from these stars \citep[e.g.][]{Kruijssen2019}. It is reasonable to expect that intermediate-age stellar populations could exist in metal-poor galaxies although the best-fitting $C_{\rm bc}\sim 1$ at low metallicity suggest the contribution from intermediate-age populations may be negligible.  
Rather, star formation across a galaxy can reasonably be seen as a constant process over the most recent 500\,Myr, possibly independent of the metallicity of the ISM in the galaxy and thus giving metallicity dependence of $\nu\sim 0$, instead of the obtained best-fitting $\nu=-3.9$, which is quite steep.  We argue that the best-fitting  $C_{\rm bc}$ and $\nu$ at low metallicity should not be thought of a measure of the UV radiation fraction from within BCs; instead it may encode physics that impacts what fraction of UV light will be subject to BC attenuation. Alternatively, the expelling of gas and dust driven by stellar feedback might become inefficient  in metal-poor BCs  and the expelled clouds could remain clumpy and act similarly to BCs in attenuating the UV light.  We notice that at low metallicity these model parameters might be biased by our assumption of $\hat{R}=\hat{H}$ (see Section~\ref{sec_5.4} for more details).

The best-fit star-to-diffuse dust radius ratio $\hat{R}$ is $\sim$0.53 in our model, which suggests that the stellar disc has a smaller radius than diffuse dust. The star formation often takes place in more central regions that feature a higher dust/gas density, while the diffuse dust extends further into the galaxy outskirts \citep{Kennicutt2012, Smith2016,Casasola2017}.  We notice that the observed dust emission consists of both dense/BC and diffuse dust. To compare with observations, we calculate the star-to-total dust radius ratio $\hat{R}^{\rm tot}$ based on $\hat{R}$ and $F_{\rm bc}$ (see Appendix~\ref{sec_a}). We can see that the star-to-total dust radius $\hat{R}^{\rm tot}$ decreases with decreasing metallicity. In the case of extremely metal-poor environments (below the metallicity coverage of our sample), $F_{\rm bc}$ is expected to be close to unity and the `total dust' is dominated by BC dust, which traces the stars.  In other words, BC dust and total dust discs are the same one, i.e, $\hat{R}^{\rm tot}=1$. Conversely, in the high-metallicity limit, $F_{\rm bc}$ approaches zero, diffuse dust plays a dominant role and therefore $\hat{R}^{\rm tot}=\hat{R}=0.52$. By examining the radial distribution of dust, stars, gas, and SFR in a sub-sample of 18 face-on spiral galaxies extracted from the DustPedia sample, \citet{Casasola2017} found the average $\hat{R}^{\rm tot}=R_{\rm SFR}/R_{\rm dust}^{\rm tot}$ is $\sim$0.57 (see their table~7). We found the average metallicity of their sample is 8.6 and a consistent $\hat{R}^{\rm tot}$ of 0.53 is predicted in our model at that metallicity.\footnote{Here the gas-phase metallicity is taken from \citet{De Vis2019} with a consistent PP04-N2 calibration.} We provide a {\sc PHTHON} package {\sc IRX\_TAU\_TOT} to calculate the total dust optical depth of a galaxy with given metallicity and best-fitting geometry parameters.\footnote{\url{https://github.com/LvZF/irx\_tau\_tot/}}

\begin{figure}
\centering
\includegraphics[width=\columnwidth]{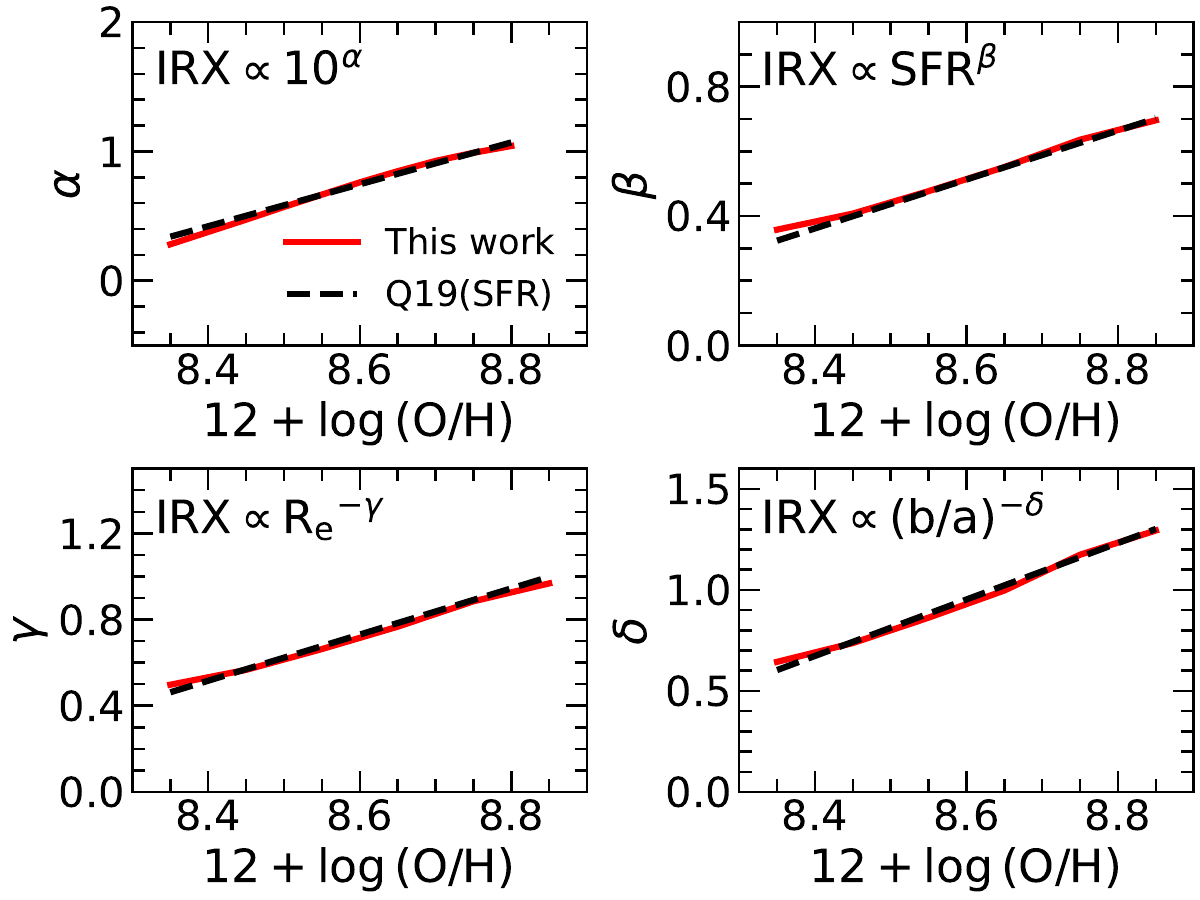}
\caption{The power-law slope of the universal IRX relation $IRX=10^\alpha SFR^{\beta} R_{\rm e}^{-\gamma} (b/a)^{-\delta}$ as a function of metallicity (dashed lines).  Results from our best-fitting two-component geometry model are overplotted as solid lines.}
\label{fig:fig4}
\end{figure}

\subsection{Comparing with the Universal IRX Relation presented in \citetalias{Qin2019a}} \label{sec4.3}

The main goal of this paper is to understand the physical origins of the universal IRX relation presented in \citetalias{Qin2019a}, particularly the systematic flattening of power slopes towards lower metallicity. 

Given a set of best-fitting model parameters, we have IRX ($Z$, SFR, $R_{\rm e}$, $b/a$). Then the IRX--Z relation can be predicted with our model at given SFR, $R_{\rm e}$ and $b/a$. As shown in \citetalias{Qin2019a}, besides the metallicity term, the power-law indices of SFR, $R_{\rm e}$ and $b/a$ in the power-law relations also contain a dependence on metallicity. One way to eliminate the metallicity dependence of other parameters and obtain a `pure' IRX--Z relation is by normalising SFR, $R_{\rm e}$ and $b/a$ to unity. For consistency, we apply the same normalisation of SFR,$R_{\rm e}$ and $b/a$ to unity in our model. We compare our thus obtained model-predicted IRX--Z relation against the relation given by \citetalias{Qin2019a} (who similarly normalised SFR, $R_{\rm e}$ and $b/a$ to unity) in the first panel of Fig.~\ref{fig:fig4}. We can see that our model perfectly reproduces the IRX-Z relation given by \citetalias{Qin2019a}. A power-law relationship adequately describes the observed IRX--Z relation.

Similarly, the SFR-IRX relation can be obtained from the best-fitting model by fixing metallicity, $R_{\rm e}$ and $b/a$. We notice that the slope of the empirical SFR--IRX relation is not constant but decreases with decreasing SFR.\footnote{Keeping in mind that the $\Sigma_{\rm dust}$ in our model is determined by the combination of $Z$, SFR, $R_{\rm e}$ and $b/a$ in the forms of power-laws.} One way to reduce this to a single measure of slope is adopting the slope of the line tangent to the curve at some representative SFR. We divide our sample galaxies into continuous metallicity bins with a constant bin width of 0.1\,dex. In each metallicity bin, we calculate the average metallicity, SFR, $R_{\rm e}$ and $b/a$. With these averaged values, we obtain a model-predicted SFR-IRX relation and then measure the slope of the tangent line to the curve at the average SFR (i.e., $\beta$). Repeating the calculation to all metallicity bins, the $\beta$ as a function of metallicity is obtained (top-right panel of Fig.~\ref{fig:fig4}).  With a similar approach, the power slope of the $R_{\rm e}$--IRX relation ($\gamma$) and $b/a$--IRX relation ($\delta$) as a function of metallicity can be also determined, respectively. As shown in the remaining panels of Fig.~\ref{fig:fig4}, all the derived power-law slopes $\beta$, $\gamma$, and $\delta$ from our best-fitting two-component geometry model agree perfectly with that in \citetalias{Qin2019a} as a function of metallicity.  We notice that the dispersion of IRX residuals decreases slightly from 0.188\,dex in the empirical power-law fitting to 0.185\,dex in the geometry model fitting. That is, the IRX relations are described slightly better by our geometry model than by the power-law forms. The latter should be regarded as an approximate description over a certain dynamic range of observed parameters.

\begin{figure*}
\centering
\includegraphics[width=0.475\textwidth, trim=-20 -5 0 0]{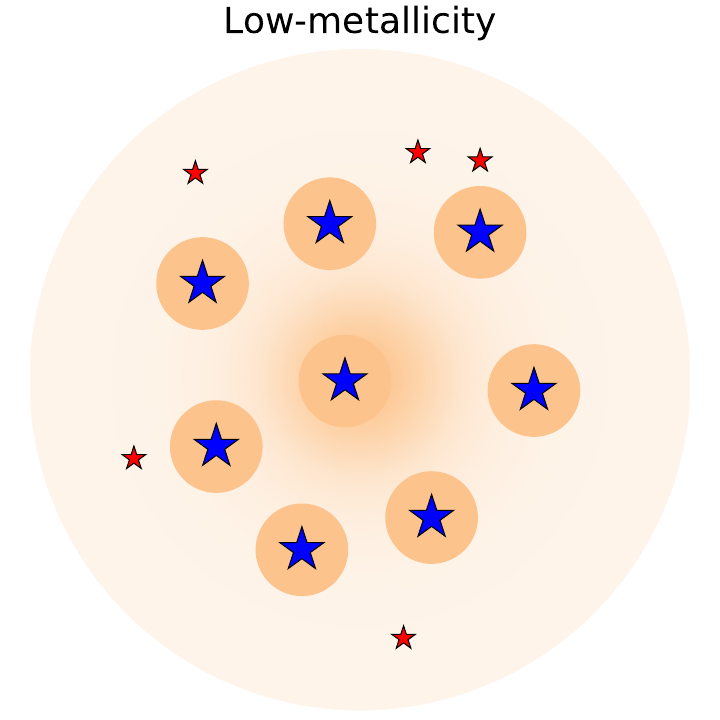}
\includegraphics[width=0.475\textwidth, trim=-20 -5 0 0]{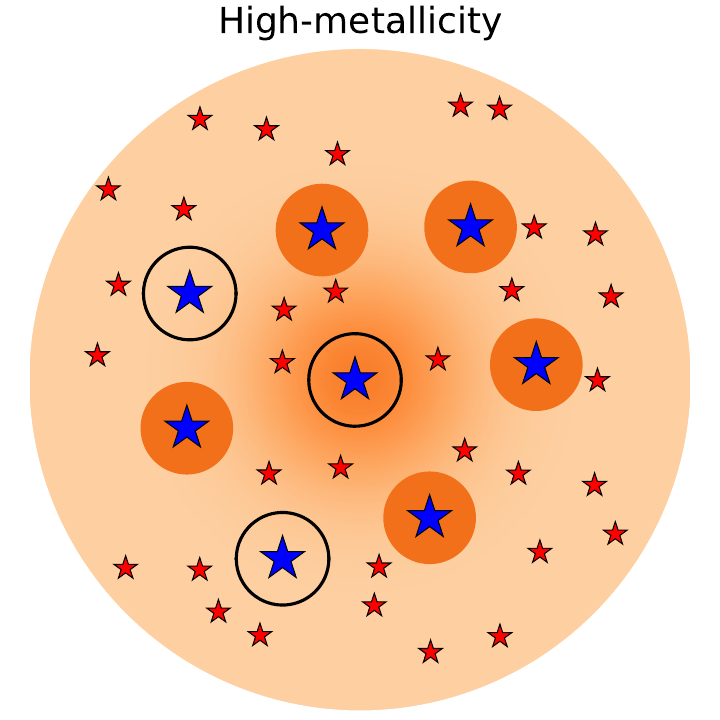}
\includegraphics[width=0.475\textwidth]{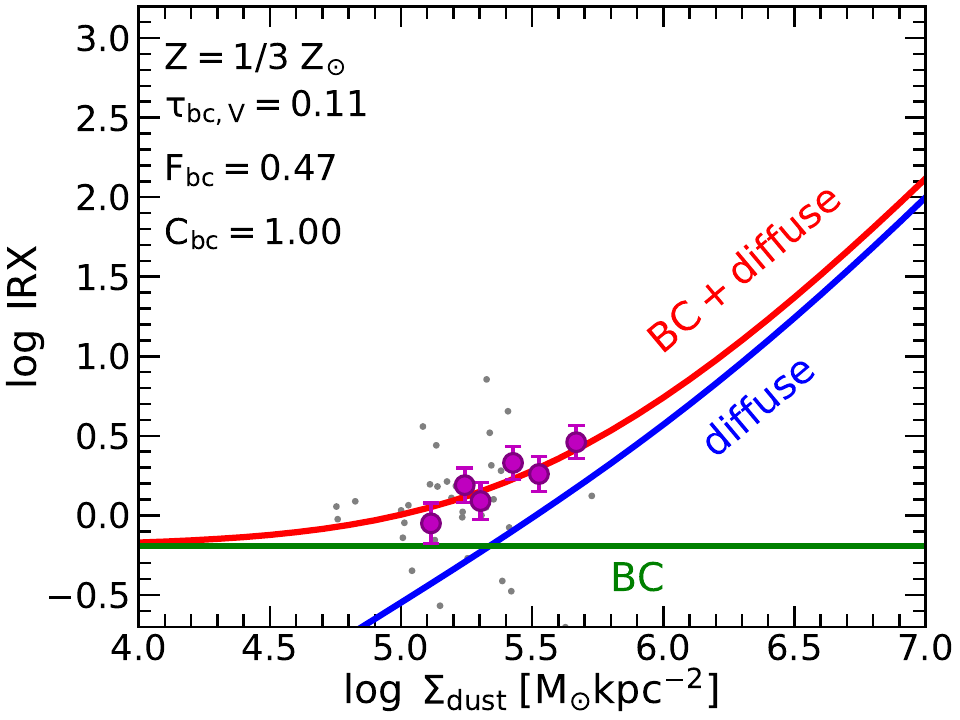}
\includegraphics[width=0.475\textwidth]{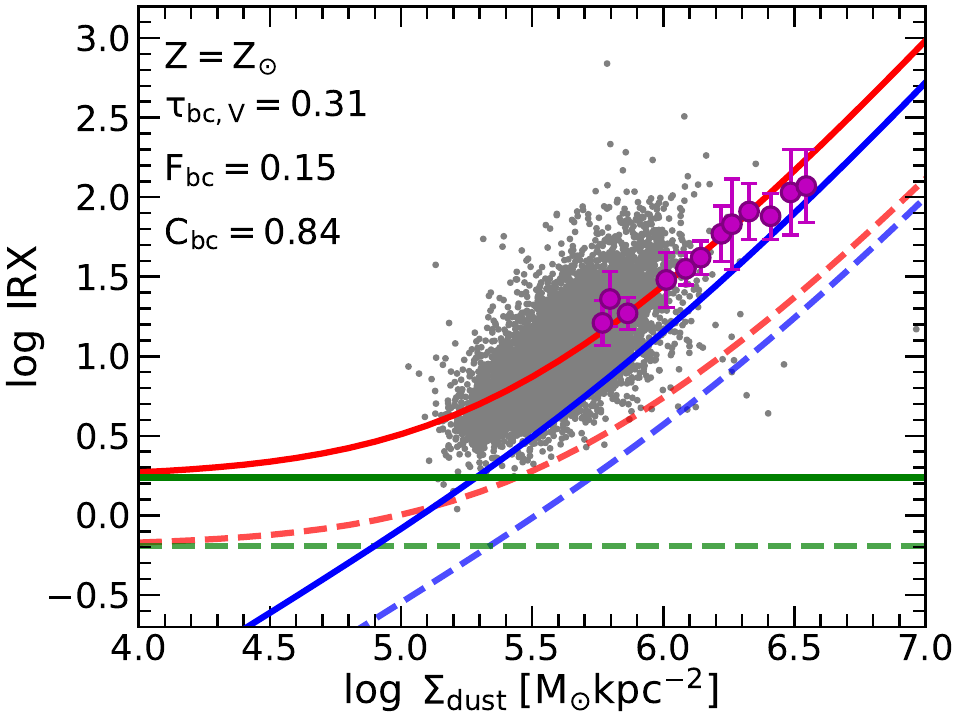}
\caption{{\it Top:} Schematic diagrams of the two-component dust geometry model in galaxies at low (top-left) and high (top-right) metallicities. The smooth background with density gradient represents the distribution of diffuse dust, while the discrete filled clumps represent the dense clouds and open circles represent the dispersed clouds due to stellar feedback from newly-born stars. The blue stars represent the young and intermediate-age stellar populations with UV emission, and red stars represent old stellar populations. {\it Bottom:} IRX as a function of $\Sigma_{\rm dust}$ at low (bottom-left) and high (bottom-right) metallicity. The blue and green lines represent the model-predicted dust attenuation caused by diffuse and dense (BC) dust, respectively, while the red curve represents the total dust attenuation. The grey points are our local SFGs and the magenta circles with error bars represent the high-$z$ SFGs in an average sense. Here the (projected) dust surface density is calculated from metallicity, SFR, $R_{\rm e}$, and $b/a$ according to Equation~\ref{eq_sigdust_proj}.} 
\label{fig:fig5}
\end{figure*}

\section{Discussion} \label{sec5}

\subsection{Constant $\tau_{\rm bc}$ at a given metallicity}\label{sec5.1}

While the IRX caused by diffuse dust monotonously increases with $\Sigma_{\rm dust}$, adding a BC component with constant optical depth can be very effective in flattening the slope of IRX relations as shown in Fig.~\ref{fig:fig1}. In other words, the key point to capture the systematic change in IRX relation slopes is that we assume the optical depth of BCs is uniquely set by metallicity but is independent of other galaxy properties, such as SFR, $R_{\rm e}$ and b/a. It is natural to question whether the average $\tau_{\rm bc}$ may depend on other galaxy properties than metallicity instead.

A galaxy consists of numerous (of the order of hundreds to millions) BCs, that follow certain distributions of ages and initial masses (determined by the star-forming process). If these distributions are universal, then the average optical depth of the BCs should be similar in each galaxy. In contrast, for example, if the BC mass function and stellar IMF depend on the global properties of galaxies, such as their SFR or gas surface density, then one might expect that the average dust opacity is correlated with such global properties. \citet{Lu2023} argued that it is reasonable to assume the optical depth of BCs (they call it `clumps') is only decided by the metallicity or dust-to-gas ratio, since the average electron density and size distribution of H\,{\small II} regions (associated with BCs) are similar for all types of local SFGs \citep{Oey2003, Liu2013, Santoro2022}. However, \citet{Santoro2022} found that the power-law slope of the H$\alpha$ luminosity function of extragalactic H\,{\small II} regions decreases with the galaxy star formation rate surface density $\Sigma_{\rm SFR}$.  This would correspond to an enhanced clustering of young stars at high gas surface densities. If the star formation efficiency is similar among the BCs, the higher SFR surface density galaxies would then have more massive BCs on average. 

We note that more massive BCs do not necessarily have a higher column density or optical depth.  The masses and sizes of giant molecular clouds (GMC) in our Milky Way are found to follow a relation of $M_{\rm GMC}\propto R^2$, i.e., a constant surface/column density \citep{Larson1981,Lombardi2010,Ballesteros-Paredes2012,Chen2020,Lada2020}. Mapping CO emission at the cloud-scale resolution for ten nearby galaxies, \citet{Rosolowsky2021} likewise found the extragalactic GMCs to follow a mass-size relation of $M_{\rm GMC}\propto R^2$. Using the data of average $\Sigma_{\rm GMC}$ of nearby galaxies given by \citet{Rosolowsky2021}, we find that, within uncertainties, the average $\Sigma_{\rm GMC}$ changes little with either global SFR or gas surface density. In brief, there is no evidence that the average $\tau_{\rm bc}$ of nearby galaxies should be correlated with the galaxy properties, particularly the SFR or gas surface density. It is a good approximation that the optical depth of BCs is only determined by metallicity without additional dependencies on other galaxy properties \citep[see also][]{Lu2023}.  Exceptions may occur in the more extreme environments of local ultra-luminous infrared galaxies (ULIRGs, \citealt{Krumholz2005}) or higher-redshift galaxies which are substantially more gas-rich and surface dense in a galaxy-averaged sense, and also feature enhanced local densities \citep{Davies2021}.

\subsection{Physical origins of the universal IRX relation} \label{sec5.2}

One main goal in this work is to explore the physical origins of the universal IRX relation, particularly the systematic changes in the slopes of IRX as functions of SFR, $R_{\rm e}$, and $b/a$ with metallicity. To do so, we built up a new `two-component' dust geometry model with flexible parameters and performed a model fitting to the observed data.  Our model fitting shows consistent results with the power-law fitting presented in \citetalias{Qin2019a}, including the systematic changes in the power-law slope of IRX with metallicity, SFR, $R_{\rm e}$ and $b/a$ (see Fig.~\ref{fig:fig4}).  We present a schematic diagram in Fig.~\ref{fig:fig5} to illustrate our result of how the dust geometry evolves with metallicity.

At first, the BC attenuation is assumed to be constant (only determined by metallicity) while the diffuse dust attenuation increases with $\Sigma_{\rm dust}$ (see bottom panels).  The latter is determined by the combination of $Z$, SFR, $R_{\rm e}$, and $b/a$ in the forms of power-laws. The changes in the slopes of the total $\Sigma_{\rm dust}$--IRX relation are controlled by the competition of the BC and diffuse dust attenuation, which is regulated by the changes in metallicity. At low metallicity, there is less diffuse dust and the BC attenuation dominates. In this case, one can expect that the total dust attenuation is only controlled by the gas-phase metallicity (changing dust-to-gas ratio) and independent of either SFR, disc size or inclination. This is why we see a flat $\Sigma_{\rm dust}$ (SFR, $R_{\rm e}$, $b/a$)--IRX relation at lower metallicity in the bottom-left panel of Fig.~\ref{fig:fig5}. 

At high metallicity, there is a large amount of diffuse dust, both due to an increasing total amount of dust and a higher diffuse dust fraction. A fraction of UV stars lose their surrounding BCs due to the stellar feedback (the empty circle), and the attenuation caused by the BCs will be somehow reduced.  Although the BC dust opacity moderately increases due to the increased dust-to-gas ratio at higher metallicity, the diffuse dust attenuation still plays a dominant role. The total dust attenuation now becomes sensitive to the changes in global dust surface density that are decided by SFR, metallicity, size and inclination.  This is why we see a steep $\Sigma_{\rm dust}$(SFR, $R_{\rm e}$, b/a)--IRX relation at high metallicity in the bottom-right panel of Fig.~\ref{fig:fig5}. 

Unlike SFR, $R_{\rm e}$, and $b/a$, the variations in galaxy metallicity do not only change the galaxy-averaged (projected) dust surface density through the associated dust-to-gas ratio, but also alter the dust geometry in galaxies. Low-metallicity galaxies have a lower dust surface density and a BC-dominated or clumpier dust geometry. The diffuse starlight embedded in a clumpy geometry will more easily escape and BC clouds themselves feature a lower opacity, altogether leading to a smaller net attenuation.  In other words, the positive Z--IRX relation in observations is shaped by the combination of a positive Z--$\Sigma_{\rm dust}$ relation as well as an anti-correlation between metallicity and clumpiness of the geometry.

While galaxy stellar mass is a crucial physical parameter governing the galaxy formation process, gas-phase metallicity directly measures the gaseous metal abundance and indirectly probes the dust abundance of the ISM, where star formation and dust obscuration are intimately connected. Therefore, a measure of dust attenuation such as IRX is more closely linked to the specific property of gas-phase metallicity than the absolute amount of stellar mass. As mentioned in Section~\ref{sec3.1}, the universal IRX relation involves SFR, $R_{\rm e}$, $Z$/Z$_\odot$, and $b/a$, but not $M_\ast$. The dependence on $b/a$ is solely attributed to the inclination effect, while SFR, $R_{\rm e}$, and $Z$/Z$_\odot$ are all related to the ISM of SFGs. Here, $R_{\rm e}$ is measured from the stellar light profile but is used as a proxy for the spatial distribution of the ISM and young stellar populations (see \citetalias{Qin2019a} for more discussion). It is important to emphasize that the dust attenuation parameter IRX is derived from the fraction between the absorbed and unabsorbed bolometric luminosity of the young and intermediate-age stellar populations that are tightly associated with the ISM, rather than the older stars that dominate the stellar mass in SFGs.  The geometry parameters in our model, such as the star-to-total dust scale-length ratio $R^{\rm tot}$, are set to be related to $R_{\rm e}$ of a galaxy and thus are connected with stellar mass. Our model fitting results show that $R^{\rm tot}$ moderately decreases with metallicity. Given that metallicity is correlated with stellar mass, the best-fitting $R^{\rm tot}$ as a function of metallicity might also reflect its dependence on stellar mass. For general populations of disc-dominated SFGs, the ISM is globally associated with older stellar populations in terms of its geometry. In conclusion, our study shows that our model parameters as a function of metallicity are able to unveil the processes that regulate the universal IRX relation.

\begin{figure}
\centering
\includegraphics[width=0.86\columnwidth]{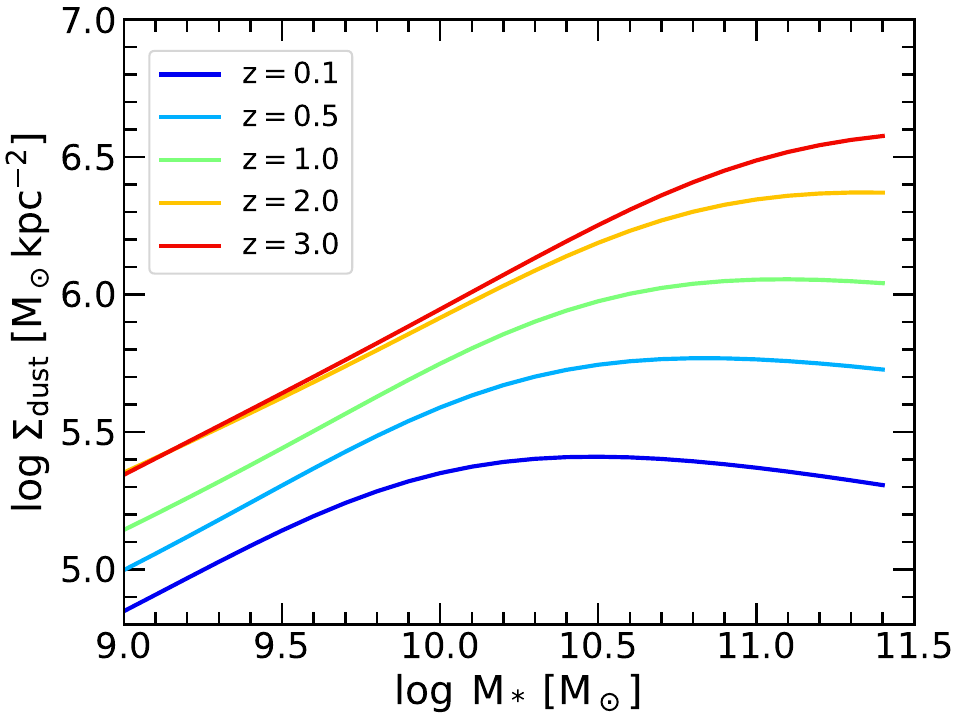}
\includegraphics[width=0.86\columnwidth]{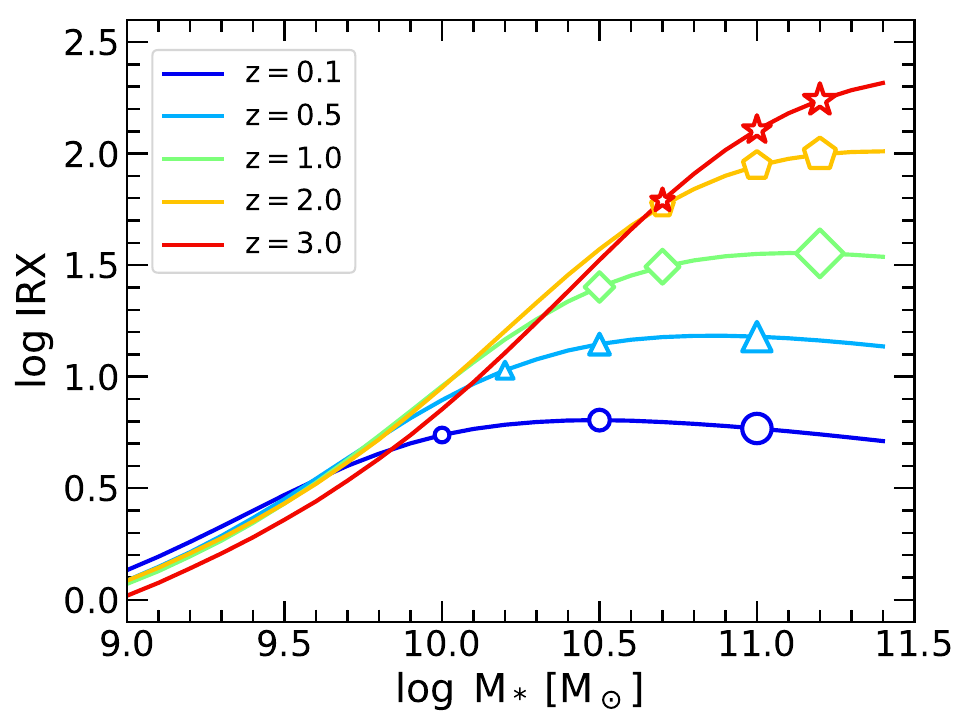}
\includegraphics[width=0.86\columnwidth]{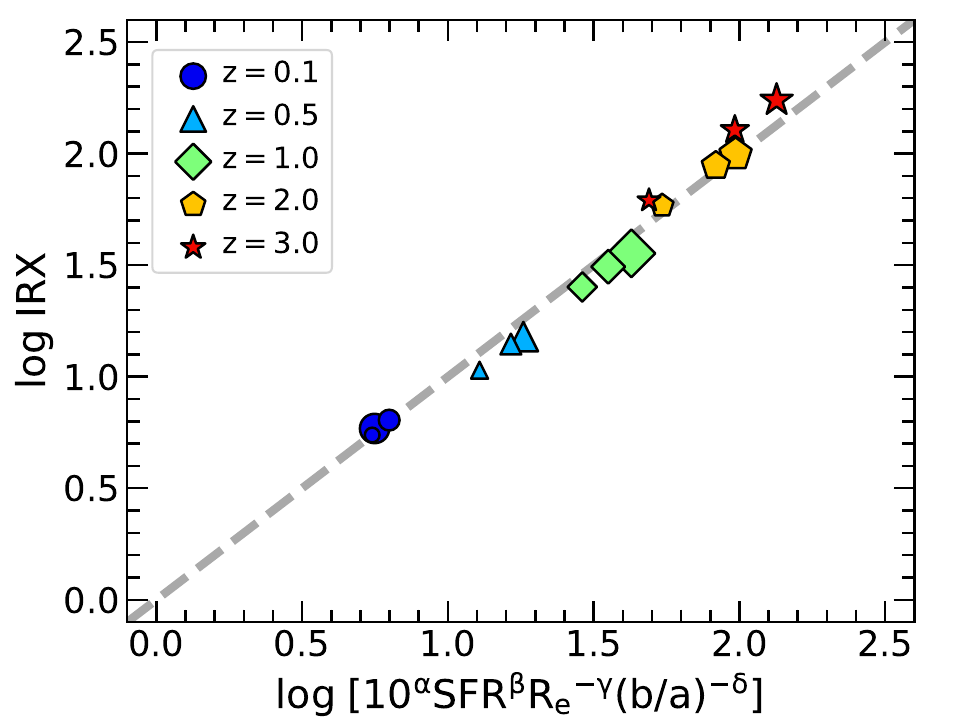}
\caption{{\it Top}: the predicted $\Sigma_{\rm dust}$ as a function of stellar mass at different redshifts over $0<z<3$. The prediction is made with our two-component star-dust geometry model with the best-fitting parameters to the local SFGs  in combination with the empirical  mass--SFR, mass--metallicity, mass--size scaling  relations, the K-S Law and the Z--DGR relation from the literature.  {\it Middle}: The predicted IRX as a function of stellar mass for redshifts $0<z<3$, obtained by pairing our dust-star geometry model with the aforementioned empirical scaling relations. {\it Bottom}: Solid symbols show the predicted IRX from our star-dust geometry model, contrasted to the IRX predicted by the universal IRX relation (Equation~\ref{eq:eq4}).  The open and solid symbols of the same size and shape correspond to the same galaxy subpopulations. Clearly, the observed mass--IRX relations given in \protect\cite{Whitaker2014} can be well reproduced in the sense that IRX mildly evolves with redshift at the high-mass end but shows no evolution at low- or intermediate stellar masses. When replacing $M_\ast$  with the predicted IRX by the universal IRX relation, the scattered data points (the open symbols) at the high-mass end in the middle panel redistribute to form a tight relation (the solid symbols) in the bottom panel. }
\label{fig:fig6}
\end{figure}

\subsection{Explaining the no evolution of mass--attenuation relation} \label{sec5.3}

Recent studies found that the relation between dust attenuation and stellar mass evolves mildly with cosmic time up to $z\sim3$ \citep[e.g.][]{Price2014, Whitaker2017, McLure2018, Shapley2022, Shapley2023, Zhang2023}. However, it introduces a `non-evolution puzzle' since either the SFR, ISM content, size, and metallicity evolve significantly with redshift at fixed stellar mass. Quantitatively, \citet{Shapley2022} found the dust surface density (estimated either via direct or indirect methods) to increase by a factor of $\sim$4 from $z=0$ to $z=2.3$ at given $M_\ast$ while the dust attenuation remains unchanged. This means that the dust attenuation per unit of $\Sigma_{\rm dust}$ (hereafter `dust attenuation efficiency') should decrease towards higher redshifts. They propose several potential origins at $z\sim2.3$: 1) a smaller dust-to-metal ratio (steeper Z--DGR relation); 2) a dust distribution that is relatively more extended compared to the stars; 3) clumpier dust distributions; and 4) a lower dust mass extinction coefficient $\kappa_{\lambda}$.

Of them, 2) is firstly excluded since high-$z$ galaxies usually have smaller far-IR sizes than their local counterparts \citep[e.g.][]{Fujimoto2017,Tadaki2020,Gomez-Guijarro2022}.\footnote{\citet{Zhang2023} note that a low galaxy-integrated attenuation can also be achieved by adopting a more compact (rather than a more extended) distribution of dust relative to the stars, thus reducing the column to all but the most central stars.  They evaluate however that the factor by which dust discs ought to be smaller is unrealistically large in such a scenario, and therefore favour the clumpy scenario too.} In our geometry model, both the Z--DGR relation and $\kappa$ are fixed, i.e., 1) and 4) are not included in our model. In other words, through the modelling study in this work, only the clumpier geometry at high-$z$ is favoured. High-$z$ galaxies usually have lower metallicity than local galaxies, and they are expected to have a clumpier geometry as shown in Fig.~\ref{fig:fig5}. In this case, more dust is in the form of BCs rather than smoothly distributed in the ISM. The decrease of diffuse dust causes a decrease in dust attenuation, while the increasing BC dust only increases the number of BCs but does not affect the global dust attenuation. On the other hand, if the BC covering fraction is not 100\%, the more clumpy geometry means that a fraction of starlight will escape directly from the galaxy without encountering clumps and suffers little dust attenuation. Both effects lead to a smaller dust attenuation at a given dust surface density.  The interpretation in terms of a more clumpy (i.e. BC-dominated) dusty geometry is also in line with the reduced (or even absent) inclination dependence of attenuation in high-z SFGs as observed out to cosmic noon by \citet{Lorenz2023} and \citet{Zhang2023}, and beyond cosmic noon by \citet{Gomez-Guijarro2023}.

Motivated by the universality of our IRX relation, we predict the mass--$\Sigma_{\rm dust}$ relation and mass--attenuation relation at different redshifts based on the best-fitting model as well as the evolution of galaxy scaling relations, including the mass--SFR relation \citep{Popesso2023}, the mass--metallicity relation \citep{Sanders2023}, and the mass--size relation \citep{van der Wel2014a}. The results are shown in Fig.~\ref{fig:fig6}. We remind the reader that such predictions are based on the assumption that all of the K--S Law, Z--DGR relation, $R_{\rm d}$--$R_\star$ relation and metallicity-dependence of dust geometry changes only mildly across cosmic time. We can see that, although a significant evolution of $\Sigma_{\rm dust}$ at given $M_\ast$ is present, the dust attenuation evolves mildly at low and intermediate masses. This indicates that a clumpy geometry at high-$z$ is sufficient to explain the puzzle of significant evolution in $\Sigma_{\rm dust}$ but no evolution in dust attenuation \citep[see also][]{Zhang2023}. Besides,  our model predicts a moderate evolution of dust attenuation at $M_\ast>10^{10.3}$\,M$_\odot$. This is consistent with \citet{Whitaker2014} who found the IRX evolves mildly at $M_\ast<10^{10.5}$\,M$_\odot$ but evolves moderately at higher masses. However, \citet{Shapley2022} did not find a moderate evolution of dust attenuation traced by the Balmer Decrement at the high-mass end. We notice that the sample used for \citet{Shapley2022} is not `very' massive, and the majority of galaxies have $M_\ast$ less than $\sim10^{10.5}$\,M$_\odot$ \citep[see also][]{Shapley2023}. Another possible reason might be attributed to the difference between the Balmer decrement and IRX indicators. The former traces nebular emission attenuation, while the latter traces the stellar emission attenuation \citep{Qin2019b}. More efforts should be made to give a self-consistent explanation. Such an analysis, however, goes beyond the scope of this work. 

\begin{figure}
\centering
\includegraphics[width=\columnwidth]{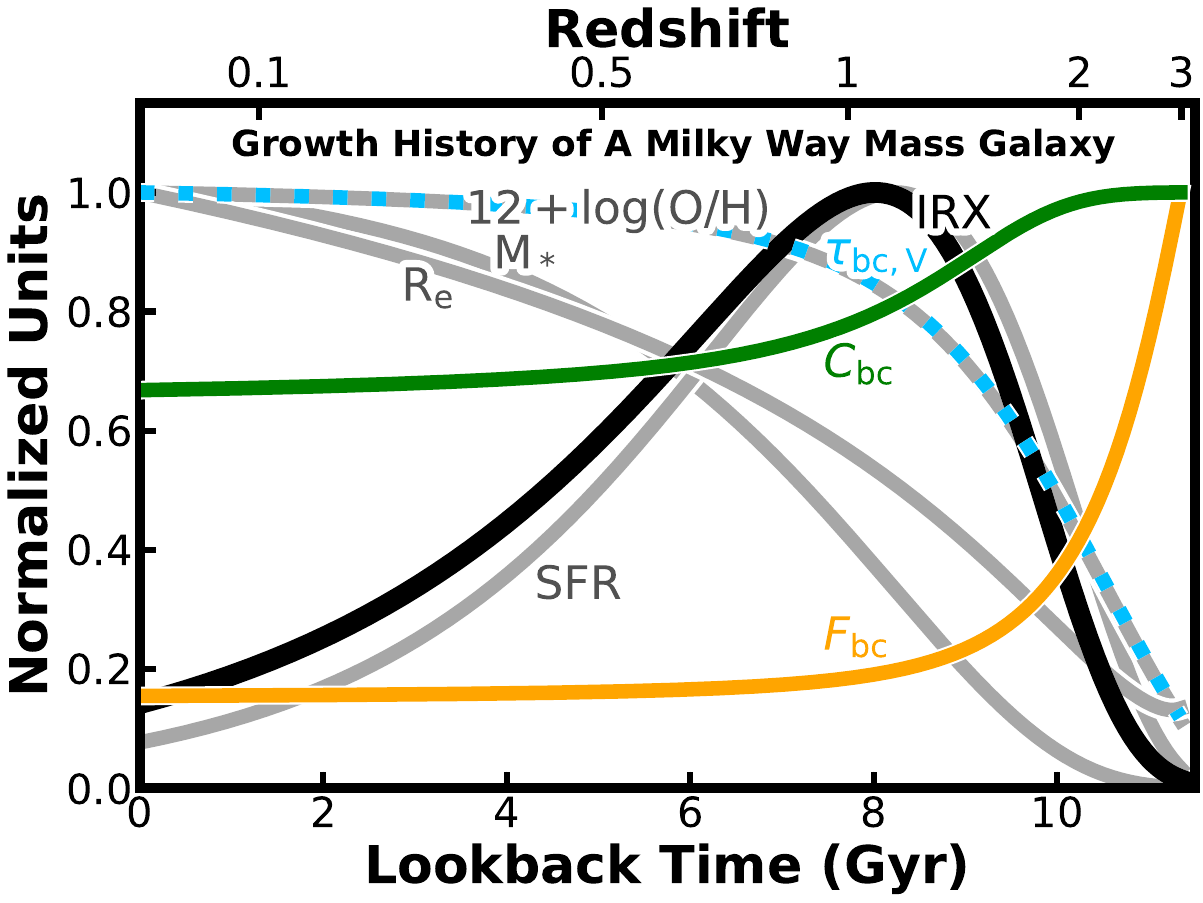}
\caption{The growth history of a Milky-Way like galaxy in given parameters/quantities: star formation rate (SFR), stellar mass ($M_\ast$), half-light radius ($R_{\rm e}$), and gas-phase metallicity [12+log(O/H)] adopted from \protect\citet{ForsterSchreiber2020}. The predicted IRX (black) and the best-fitting parameters $\tau_{\rm bc,V}$ (cyan), $F_{\rm bc}$ (orange) and $C_{\rm bc}$ (green) from our two-component star-dust geometry model are shown for comparison.}
\label{fig:fig7}
\end{figure}

Given that both distant and nearby SFGs conform to the universal IRX relation, it is intriguing to explore the evolutionary trajectory of a galaxy from high redshift to the present day. Extracting such an evolutionary track from the IRX-mass-redshift relation depicted in the middle panel of Fig.~\ref{fig:fig6} requires connecting data points representing low-mass SFGs at $z=3$ with more massive SFGs at lower redshifts.  Following the methodology outlined by \citet{ForsterSchreiber2020} to retrieve the evolutionary paths of a Milky-Way-like galaxy from $z=3$ to the present day, we show in Fig.~\ref{fig:fig7} their derived histories of star formation rate (SFR), stellar mass ($M_\ast$), effective radius ($R_{\rm e}$), and metallicity ($Z/$Z$_\odot$). Additionally, for comparison, we present the corresponding best-fitting parameters ($\tau_{\rm bc,V}$, $F_{\rm bc}$ and $C_{\rm bc}$) of our two-component star-dust geometry model, as well as the model-predicted IRX. It can be observed that, in general, the IRX curve initially follows the SFR curve, with deviations likely resulting from metallicity- and structure-related processes. A striking transition occurs roughly around $z\sim 1.5$ from the metal-poor and compact progenitor to the metal-rich and extended disky galaxy. 

\subsection{Caveats}\label{sec_5.4}

For simplicity, we have utilized the Calzetti Law as the fixed galaxy dust attenuation curve in our modelling. Additionally, we have assumed that the star-to-diffuse dust disc scale-length and scale-height ratios follow each other as $\hat{R} =\hat{H}$. The Calzetti Law has been successfully applied in our modelling of the universal IRX relation for the overall population of SFGs. It is important to note that the dust attenuation curve may differ for various galaxy populations. For example, the Milky Way, Large Magellanic Cloud, and Small Magellanic Cloud exhibit different curves. Previous studies have reported variations in the slope of the dust attenuation curves for different galaxy samples, which can be either steeper than the Calzetti Law \citep[e.g.][]{Shivaei2020} or shallower \citep[e.g.][]{Salim2020}. The slope of the dust attenuation curves and dust attenuation were reported to be anti-correlated \citep{Salim2020}, likely influenced by degeneracies in galaxy spectral energy distribution (SED) fitting \citep{Qin2022}. Furthermore, the inclination of galaxies has been found to affect the shape of the observed attenuation curve, with edge-on galaxies exhibiting a grayer attenuation curve compared to face-on galaxies \citep[see e.g., Appendix of][]{Zhang2023}. These findings suggest that incorporating flexibility to the dust attenuation curve could further improve the realism with which observations are reproduced by our modelling, and could provide insights into the factors influencing the slope of the effective dust attenuation curve.

A greyer curve, as reported for highly-inclined (i.e., edge-on) SFGs \citep[e.g.][]{Lu2022}, may result in a lower $\tau_{\rm FUV}$ at a given IRX, as a relatively larger portion of the dust-absorbed energy that is reprocessed into the IR would originate from stellar emission at wavelengths longward of the FUV regime. In such cases, the use of the Calzetti Law in our modelling could lead to an overestimation of $\tau_{\rm FUV}$. However, we anticipate the model parameters best-fitting the observed IRX, metallicity, SFR, and $R_{\rm e}$ to remain largely unchanged.  We note that the highly-inclined galaxies represent only a small fraction (2.5\% for SFGs with $b/a<0.2$) of our sample of SFGs, leaving the results of our model fitting unaffected. We acknowledge that our model does not have the capability to determine the slope of the dust attenuation curve as enabled for example by the radiative transfer approaches employed by \citet{Zhang2023}, but note that this also falls beyond the scope of this work.

Our assumption that $\hat{R} =\hat{H}$, meaning that the stellar disc and the dust disc mirror each other in shape with a scaling factor, ensures that the axial ratios of the stellar and dust components are identical. This assumption simplifies the calculation of inclination-related effects and is a crucial parameter for our two-component star-dust geometry model. Our results demonstrate that this assumption holds well on a global level. However, in real galaxies, the values of $\hat{R}$ and $\hat{H}$ can vary within a range and may not necessarily mirror each other on an individual basis. For example, massive disc galaxies like NGC\,891 often exhibit a thick stellar disc and a thin dust disc, resulting in an apparent dust lane when viewed edge-on. Such variations in $\hat{R}$ and $\hat{H}$ can introduce scatter in the dust surface density ($\Sigma_{\rm dust}$) relative to star formation and, consequently, in dust attenuation.
For metal-poor galaxies, their morphologies tend to be spherical or rounder compared to regular discs, resulting in a higher $\hat{H}$ relative to $\hat{R}$. Such star-dust geometries significantly reduce the inclination-dependent effect on dust attenuation (\citetalias{Qin2019a}). Metal-poor galaxies may thus require distinct model parameters $\hat{R}$ and $\hat{H}$ compared to those determined predominantly by the metal-rich galaxies in our sample. The fixed $\hat{R}=\hat{H}$  in our model fitting may bias $C_{\rm bc}$ towards unity in order to dramatically reduce the inclination-dependent effect at low metallicity. On the other hand, the dust attenuation indicator, IRX, is primarily influenced by the radiation from young and intermediate-age stellar populations, which are closely associated with the dust disc. IRX is less sensitive to the contribution from old stellar populations. It is important to note that a comprehensive analysis of dust attenuation should consider the geometric differences between stars and dust, accounting for possible variations in both $\hat{R}$ and $\hat{H}$.

Dust attenuation in our model accounts for absorption of light and does not take into consideration the effect of scattering. Light scattering would introduce additional light for galaxies seen face-on and less for galaxies seen edge-on. We believe this effect plays a secondary role in affecting the relationship between IRX and effective dust surface density. In contrast, the orientation effect of increasing projected dust columns with increasing inclination, which can be accurately quantified, has a more significant influence on this relationship.

\section{Summary} \label{sec6}

In this work, we have developed a new two-component dust model to describe the universal dust attenuation relation in local galaxies. We fit our model to the same SFG data exploited in \citetalias{Qin2019a} using a Bayesian MCMC sampling. Our main findings are summarized as follows:

\begin{enumerate}[i)]

	\item Our model produces a good fit to the observational data, and successfully reproduces the systematic changes in IRX scaling relations over the entire metallicity range as presented in \citetalias{Qin2019a}.

	\item We give constraints on three galaxy scaling relations through model fitting, including the star-formation law of $\Sigma_{\rm SFR}\propto\Sigma_{\rm gas}^{1.84}$, the dust-stellar radius relation of $R_{\rm e,dust}\propto R_{\rm e,star}^{0.69}$, and the metallicity-dust/gas relation of $DGR\propto Z$. All of these relations are consistent with the observational studies.  
	
	\item The evolution of dust geometry as a function of gas-phase metallicity is also constrained quantitatively. As metallicity increases from 1/3 Solar to Solar, the star-to-total dust disc scale-length ratio decreases from 0.69 to 0.53, the optical depth of birth clouds increases from 0.11 to 0.34, the birth cloud mass fraction decreases from 0.42 to 0.14, and the BC covering fraction of UV light decreases from $\sim$1 to 0.75. Low-metallicity galaxies not only have a smaller dust surface density but also a clumpier geometry. 

	\item We find the variations in the slopes of IRX with SFR, $R_{\rm e}$ and $b/a$ stem from the competition between BC and diffuse dust attenuation in galaxies, which is controlled by the galaxy metallicity. When a galaxy is metal-rich, there is a large amount of diffuse dust in its ISM, and IRX increases with the global dust surface density; At low-metallicity, the birth cloud attenuation becomes important and even dominant, and then the IRX becomes insensitive to the changes of either SFR, $R_{\rm e}$ or $b/a$. 
		
\end{enumerate}

\section*{acknowledgments}

We are grateful to the anonymous referee for valuable comments and suggestions that improved this manuscript.
This work is supported by the National Key Research and Development Program of China (2023YFA1608100), the National Science Foundation of China (12073078, 12233005, 12173088 and 12033004), the science research grants from the China Manned Space Project with NO. CMS-CSST-2021-A02, CMS-CSST-2021-A04 and CMS-CSST-2021-A07, and the Jiangsu Funding Program for Excellent Postdoctoral Talents (2022ZB473). XZZ thanks  the CAS South America Centre for Astronomy (CASSACA) for the hospitality of a three-month visit. S.W. acknowledges support from the Chinese Academy of Sciences President's International Fellowship Initiative (grant no. 2022VMB0004). A.K. has been supported by the 100 talents program of Sun Yat-sen University. V.G. gratefully acknowledges support by the ANID BASAL project FB210003 and  from ANID FONDECYT Regular 1221310.

\section*{Data Availability}
The data underlying this article will be shared on reasonable request to the corresponding author.  

%

\appendix

\onecolumn

\section{The influence of the disc thickness} \label{sec_c}

\begin{figure}
\centering
\includegraphics[width=0.9\textwidth]{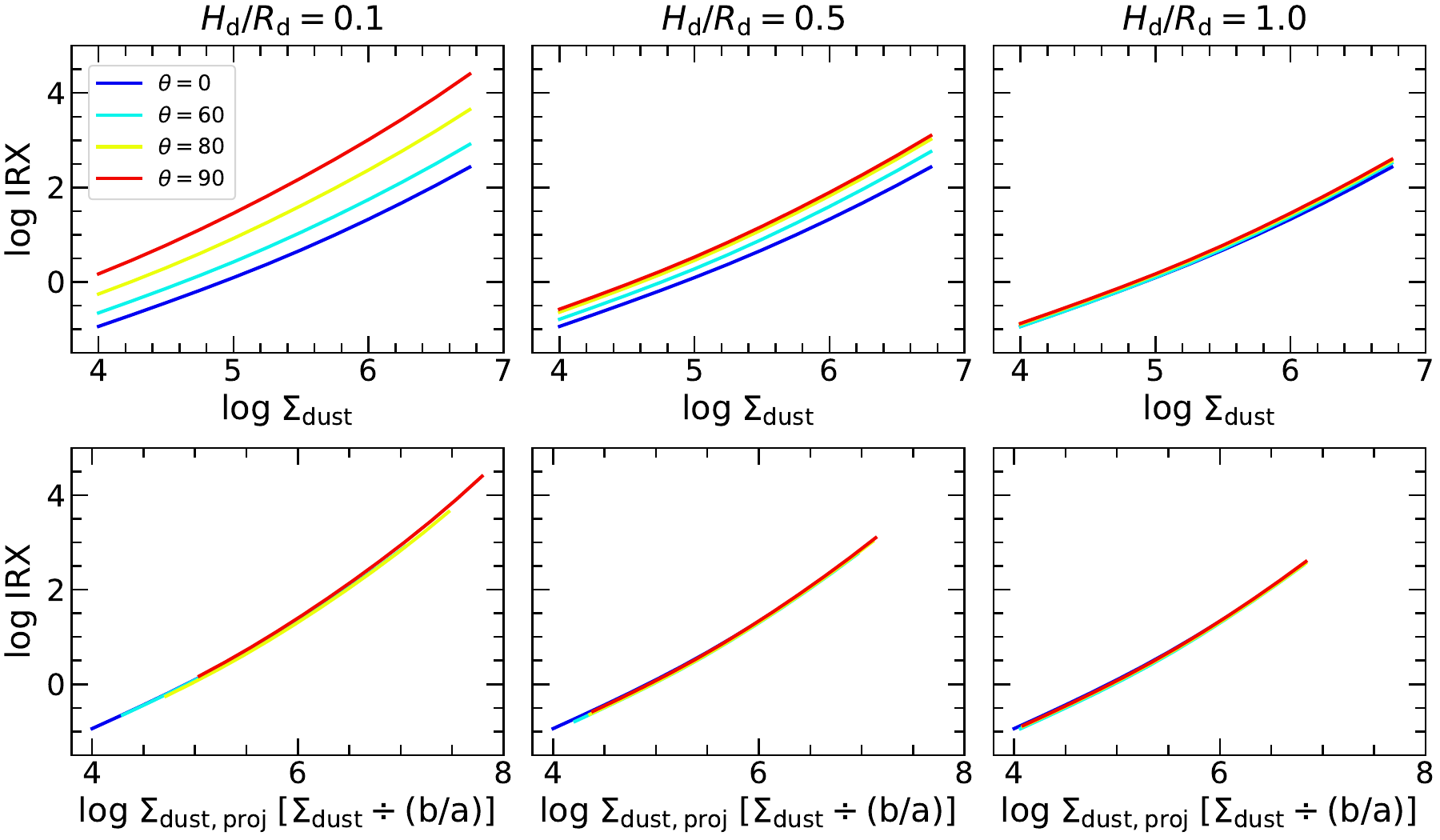}
\caption{{\it Top panels:} IRX a function of $\Sigma_{\rm dust}$ with disc thickness $H_{\rm d}/R_{\rm d}$ of 0.1 (top-left), 0.5 (top-middle), and 1 (top-right). The solid lines with blue-to-red colour represent the increasing value of inclination $\theta$. {\it Bottom panels:} similar to the top but replacing $\Sigma_{\rm dust}$ with projected dust surface density $\Sigma_{\rm dust,proj}=\Sigma_{\rm dust}\div(b/a)$. }
\label{fig:figa1}
\end{figure}

As mentioned in Section~\ref{sec3.2}, the derivation of our geometry model equations is based on the assumption that galaxies are oriented face-on. We include the axial ratio to our model by interpreting it as a simple projection effect, i.e., an edge-on galaxy (axial ratio of $b/a$) with an observed (i.e., projected) dust surface density of $\Sigma_{\rm dust}$ is in our model treated as a face-on galaxy with a dust surface density of $\Sigma_{\rm dust}\div(b/a)$. This simplified approach ignores the fact that for an inclined observer the sightline to stars at any galactocentric radius will pierce through dust located at a range of galactocentric radii, an effect that will be more pronounced for thicker discs and/or closer to edge-on viewing angles. In other words, disc thickness may bias the projection effect. Does it affect our results?

We start by considering the $\tau_{\rm rh}$ in Equation~\ref{eq_tau_rh}. At first, we assume a galaxy has a certain inclination $\theta$ with respect to the observer, defined such that $\theta = 0^\circ$ corresponds to a face-on viewing angle.  We further treat the galaxy as an axi-symmetric system.  Rewriting $\tau_{\rm rh}$ in Equation~\ref{eq_tau_rh} as $\tau_{xyh\theta}$, and introducing $l$ as the integration variable capturing distance along the line of sight, we obtain
\begin{equation}
\begin{split}
	\tau_{xyh\theta} &=\int_l^\infty \kappa \rho_0\exp\left(-\frac{\sqrt{x^2+y^2}}{R_{\rm d}}-\frac{|h|}{H_{\rm d}}\right) dl \\
			&=\frac{\kappa\Sigma_{\rm d}^{\rm diff}}{4 H_{\rm d}}\int_z^\infty \exp\left(-\frac{\sqrt{x^2+(y+(h^\prime-h)\tan\theta)^2}}{R_{\rm d}}-\frac{|h^\prime|}{H_{\rm d}}\right) \frac{dh^\prime}{\cos\theta} , 
\end{split}
\end{equation}
where $h'= \frac{h}{H_{\rm d}}$ and $\Sigma_{\rm d}^{\rm diff}$ refers to the galaxy-averaged surface density of the diffuse dust disc.
The expression for the effective optical depth due to diffuse dust (Equation~\ref{eq_tau_diff}) should likewise be rewritten as  
\begin{align}
	\tau^{\rm diff}=-\ln\int_{-\infty}^\infty dx\int_{-\infty}^{\infty}dy \int_{-\infty}^{\infty} \frac{1}{4\pi R_{\star}^2 H_{\star}}\exp\left[-\frac{\sqrt{x^2+y^2}}{R_{\star}}-\frac{|h|}{H_{\star}}-\tau_{xyh\theta}\right] dh .
\end{align}
We can see $\tau^{\rm diff}$ is a function of $\kappa$, $\Sigma_{\rm dust}$, $R_{\star}$, $H_{\star}$, $R_{\rm d}$, $H_{\rm d}$ and $\theta$. We notice that we mainly focus on the influence of relative thickness but do not care about the absolute value of the galaxy radius. Here we let $R_{\rm d}=5$\,kpc. We have checked that using another value of radius does not alter our conclusion. According to the our best-fitting model, we let $\hat{R}=\hat{H}\approx 0.5$, i.e., $R_{\star}=2.5$\,kpc, $H_{\star}=H_{\rm d}/R_{\rm d}\times2.5$\,kpc. The $\kappa$ is fixed as $0.68\times 10^{-5}$\,M$_{\odot}^{-1}\ {\rm kpc}^2$ in the $V$-band following Section~\ref{sec3.4}. Now, the optical depth of diffuse dust $\tau_{\rm diff}$ depends on the face-on $\Sigma_{\rm dust}$, thickness $H_{\rm d}/R_{\rm d}$, and inclination $\theta$.  We integrate the equations numerically to obtain the dust opacity at different inclinations. Besides, we project the three-dimensional galaxy disc onto a plane with a viewing angle of $\theta$ for a set of different thicknesses. We then use the python package \textsc{photutils}\footnote{\url{https://photutils.readthedocs.io/en/stable/index.html}} to fit the projected ellipse to obtain the axial ratio.  

The top-left panel of Fig.~\ref{fig:figa1} shows IRX as a function $\Sigma_{\rm dust}$ for a disc-like galaxy with a thickness of 0.1. We can see that at a given $\Sigma_{\rm dust}$, more inclined galaxies have higher IRX. If the thickness increases to 1 ($H_{\rm d}=R_{\rm d}$, the top-right panel), the shape of the galaxy is close to a sphere, and therefore the dust opacity becomes no longer sensitive to the changes of inclination. The bottom panels show the IRX as a function of projected dust surface density, determined by dividing $\Sigma_{\rm dust}$ by $b/a$. Surprisingly, at a given $\Sigma_{\rm dust,proj}$, the dependence of IRX on inclination disappears for different thicknesses. Although our derivation of the model formulas is based on the face-on galaxy orientation, including the axial ratio via a simple projection approach enables our results not to be affected.

\section{Determining the numerical solution $\hat{R}^{\lowercase{tot}}(\hat{R}, F_{\lowercase{bc}})$} \label{sec_a}

\begin{figure}
\centering
\includegraphics[width=0.55\textwidth]{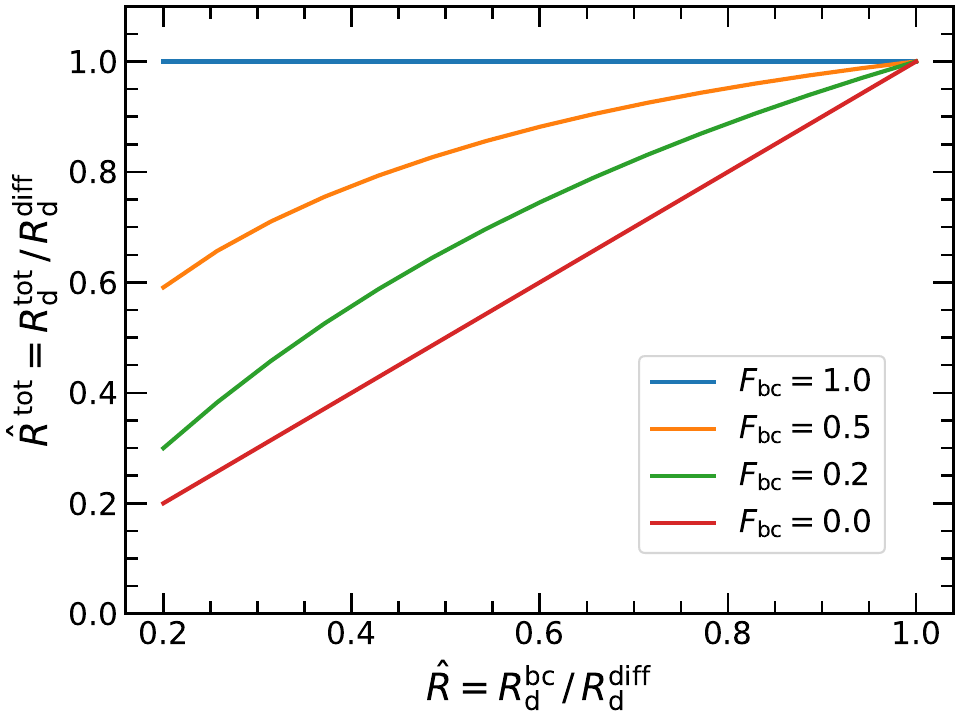}
\caption{$\hat{R}^{\rm tot}$ as a function $\hat{R}$ at different $F_{\rm bc}$. Here, $\hat{R}^{\rm tot}$ represents the size ratio of total dust and diffuse dust disc, $\hat{R}$ the size ratio of SFR (or BC) and diffuse dust disc, and $F_{\rm bc}$ the fraction of dust comprised in BCs.}
\label{fig:figb1}
\end{figure}

We here derive the relation between the size ratio of SFR and dust disc when accounting for all dust ($\hat{R}^{\rm tot}$) or only diffuse dust ($\hat{R}$).  As we will demonstrate, the relation between the two varies with changing birth cloud dominance ($F_{\rm bc}$).
Again, we use no superscript, superscript $bc$, and superscript $tot$ to denote the diffuse dust, BC dust and total dust components, respectively. We assume both BC and dust disc follow exponential profiles as   
\begin{align} \label{eq_A1}
	\rho_{\rm d}^{bc}(r)&=\frac{F_{\rm bc}}{R_{\rm d}^{\rm bc}} \exp(-r/R_{\rm d}^{\rm bc}) ,\\
	\rho_{\rm d}(r)&=\frac{1-F_{\rm bc}}{R_{\rm d}} \exp(-r/R_{\rm d}) \label{eq_A2} ,
\end{align}
and the integrated mass of BC and diffuse dust are $F_{\rm bc}$ and $1-F_{\rm bc}$, respectively (i.e., for simplicity we use a total dust mass of unity in our derivation). Then the total dust (BC+diff) profile is 
\begin{align} \label{eq_A3}
	\rho_{\rm d}^{\rm tot}(r)=\frac{F_{\rm bc}}{R_{\rm d}^{\rm bc}} \exp(-r/R_{\rm d}^{\rm bc})+\frac{1-F_{\rm bc}}{R_{\rm d}} \exp(-r/R_{\rm d}) .
\end{align}
Approximating the total profile as an exponential disc with $\rho^{\rm tot}(r)=1/R_{\rm d}^{\rm tot}\exp(-r/R_{\rm d}^{\rm tot})$ (total dust mass is unity), we then have 
\begin{align} \label{eq_A4}
	\frac{1}{R_{\rm d}^{\rm tot}} \exp(-r/R_{\rm d}^{\rm tot})=\frac{F_{\rm bc}}{R_{\rm d}^{bc}}\exp(-r/R_{\rm d}^{bc})+\frac{1-F_{\rm bc}}{R_{\rm d}}\exp(-r/R_{\rm d})	.
\end{align}

Multiplying both sides of the Equation~\ref{eq_A4} by $R_{\rm d}$ and letting  $r^\prime=r/R_{\rm d}$, then Equation~\ref{eq_A4} can be rewritten as , 
\begin{align} \label{eq_A5}
	\frac{R_{\rm d}}{R_{\rm d}^{\rm tot}} \exp(-r^\prime \frac{R_{\rm d}}{R_{\rm d}^{\rm tot}})=F_{\rm bc}\frac{R_{\rm d}}{R_{\rm d}^{\rm bc}}\exp(-r^\prime \frac{R_{\rm d}}{R_{\rm d}^{\rm bc}})	+(1-F_{\rm bc}) \exp(-r^\prime)
\end{align}
Given that the BC dust disc is associated with the UV-emitting stellar disc, we can therefore write $\hat{R}=R_\star/R_{\rm d}=R_{\rm d}^{bc}/R_{\rm d}$ and $\hat{R}^{\rm tot}=R_{\rm d}^{bc}/R_{\rm d}^{\rm tot}$.   This allows Equation~\ref{eq_A5} to be rewritten as  
\begin{align}\label{eq_A6}
	\frac{1}{\hat{R}^{\rm tot}} \exp(-r^\prime/ \hat{R}^{\rm tot})=\frac{F_{\rm bc}}{\hat{R}^{\rm tot}}\exp(-r^\prime/\hat{R}^{\rm tot})+(1-F_{\rm bc})\exp(-r^\prime).
\end{align}
For a given $F_{\rm bc}$ and $\hat{R}$, we thus have a profile for the total dust disc and the $\hat{R}^{\rm tot}$ can be obtained by fitting the profile (within $r^\prime<3$). In other words, $\hat{R}^{\rm tot}$ is a function of $\hat{R}$ and $F_{\rm bc}$. We show the numerically derived $\hat{R}^{\rm tot}$ as a function of $\hat{R}$ at different $F_{\rm bc}$ in Fig.~\ref{fig:figb1}. We can see that $\hat{R}^{\rm tot}$ increases with $\hat{R}$ and decreases with $F_{\rm bc}$. When $F_{\rm bc}=1$, i.e., there is no diffuse dust, $\hat{R}^{\rm tot}=1$; When $F_{\rm bc}=0$, i.e., there is no BC dust, $\hat{R}^{\rm tot}=\hat{R}$. It is worthwhile noting that the total dust disc can be well described by an exponential profile in most cases. In some extreme cases, e.g. at $\hat{R}<0.3$, it slightly deviates from an exponential profile. This will not affect our results since our model fitting suggests the $\hat{R}\approx0.5$.       

\section{Calibrating the total infrared luminosity from WISE mid-IR bands}\label{sec_b}

We calibrate the total infrared luminosity by making use of the high-quality data from the DustPedia survey \citep{Davies2017, Clark2018}.\footnote{\url{http://www.dustpedia.astro.noa.gr/}} The total IR luminosities are taken from \citet{Nersesian2019}, in which \textsc{CIGALE} was adopted to perform the energy balance fitting and the \textsc{THEMIS} model was used to derive the dust properties. Galaxies that have large errors on either stellar mass, SFR or IR luminosity (i.e., $>0.3$\,dex) were excluded from our analysis. Our final sample contains 324 SFGs with secure detections in multiple bands.      

\begin{figure}
\centering
\includegraphics[width=0.9\textwidth]{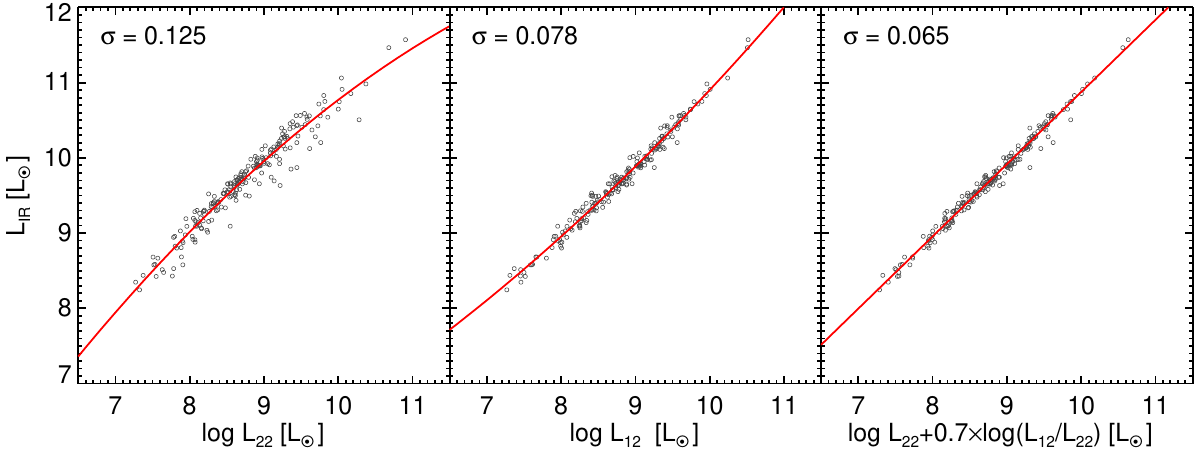}
\caption{$L_{\rm IR}$ as a function of $L_{12}$, $L_{22}$ and their combination for the DustPedia sample. The red lines depict the corresponding best-fitting relations and the 1\,$\sigma$ scatter of the sample around this relation is also given.}
\label{fig:figc1}
\end{figure}

To better sample the IR SED range and obtain reliable $L_{\rm IR}$ measurements, we require detections with a signal-to-noise ratio $S/N>3$ in at least three Herschel bands (70--500\,$\mu$m). We give a new calibration of total infrared luminosity [$L_{\rm IR}$ (8--1000\,$\mu$m)] as a function of the combination of WISE 12\,$\mu$m and 22\,$\mu$m luminosities, 
\begin{equation}
	\log L_{\rm IR}=1.19 + 0.97 \times ( \log L_{22} + 0.7 ( \log L_{12} - \log L_{22} ) ) ,
\end{equation} 
where $L_{\rm 12}$ and $L_{\rm 22}$ are the monochromatic luminosities given by $L_{12}=\nu L_{\nu}(12\,\mu\rm m)$ and $L_{22}=\nu L_{\nu}(22\,\mu\rm m)$  in units of Solar luminosity. 
Fig.~\ref{fig:figc1} demonstrates that the scatter significantly decreases if two mid-IR bands are used to constrain the total IR luminosity.

\bsp	
\label{lastpage}

\begin{thebibliography}{99}


\bibitem[\protect\citeauthoryear{Aoyama et al.}{2017}]{Aoyama2017} Aoyama S., Hou K.-C., Shimizu I., Hirashita H., Todoroki K., Choi J.-H., Nagamine K., 2017, MNRAS, 466, 105  
\bibitem[\protect\citeauthoryear{Asano et al.}{2013}]{Asano2013} Asano R.~S., Takeuchi T.~T., Hirashita H., Nozawa T., 2013, MNRAS, 432, 637  
\bibitem[\protect\citeauthoryear{Baes et al.}{2020}]{Baes2020} Baes M., et al., 2020, A\&A, 641, A119  
\bibitem[\protect\citeauthoryear{Ballesteros-Paredes, D'Alessio, \& Hartmann}{2012}]{Ballesteros-Paredes2012} Ballesteros-Paredes J., D'Alessio P., Hartmann L., 2012, MNRAS, 427, 2562  
\bibitem[\protect\citeauthoryear{Battisti, Calzetti, \& Chary}{2016}]{Battisti2016} Battisti A.~J., Calzetti D., Chary R.-R., 2016, ApJ, 818, 13  
\bibitem[\protect\citeauthoryear{Bell et al.}{2005}]{Bell2005} Bell E.~F., et al., 2005, ApJ, 625, 23  
\bibitem[\protect\citeauthoryear{Bolatto, Wolfire, \& Leroy}{2013}]{Bolatto2013} Bolatto A.~D., Wolfire M., Leroy A.~K., 2013, ARA\&A, 51, 207  
\bibitem[\protect\citeauthoryear{Calabr{\`o} et al.}{2017}]{Calabro2017} Calabr{\`o} A., et al., 2017, A\&A, 601, A95  
\bibitem[\protect\citeauthoryear{Calzetti, Kinney, \& Storchi-Bergmann}{1994}]{Calzetti1994} Calzetti D., Kinney A.~L., Storchi-Bergmann T., 1994, ApJ, 429, 582  
\bibitem[\protect\citeauthoryear{Calzetti et al.}{2000}]{Calzetti2000} Calzetti D., Armus L., Bohlin R.~C., Kinney A.~L., Koornneef J., Storchi-Bergmann T., 2000, ApJ, 533, 682  
\bibitem[\protect\citeauthoryear{Casasola et al.}{2017}]{Casasola2017} Casasola V., et al., 2017, A\&A, 605, A18  
\bibitem[\protect\citeauthoryear{Catinella et al.}{2018}]{Catinella2018} Catinella B., et al., 2018, MNRAS, 476, 875  
\bibitem[\protect\citeauthoryear{Charlot \& Fall}{2000}]{Charlot2000} Charlot S., Fall S.~M., 2000, ApJ, 539, 718  
\bibitem[\protect\citeauthoryear{Chen et al.}{2020}]{Chen2020} Chen B.-Q., et al., 2020, MNRAS, 493, 351  
\bibitem[\protect\citeauthoryear{Clark et al.}{2018}]{Clark2018} Clark C.~J.~R., et al., 2018, A\&A, 609, A37  
\bibitem[\protect\citeauthoryear{Conroy}{2013}]{Conroy2013} Conroy C., 2013, ARA\&A, 51, 393  
\bibitem[\protect\citeauthoryear{Curti et al.}{2023}]{Curti2023} Curti M., et al., 2023, arXiv, arXiv:2304.08516  
\bibitem[\protect\citeauthoryear{Dale \& Helou}{2002}]{Dale2002} Dale D.~A., Helou G., 2002, ApJ, 576, 159  
\bibitem[\protect\citeauthoryear{Davies et al.}{2017}]{Davies2017} Davies J.~I., et al., 2017, PASP, 129, 044102  
\bibitem[\protect\citeauthoryear{Davies et al.}{2021}]{Davies2021} Davies R.~L., et al., 2021, ApJ, 909, 78  
\bibitem[\protect\citeauthoryear{De Vis et al.}{2019}]{De Vis2019} De Vis P., et al., 2019, A\&A, 623, A5  
\bibitem[\protect\citeauthoryear{Draine \& Li}{2007}]{Draine2007a} Draine B.~T., Li A., 2007, ApJ, 657, 810  
\bibitem[\protect\citeauthoryear{Draine et al.}{2007}]{Draine2007b} Draine B.~T., et al., 2007, ApJ, 663, 866  
\bibitem[\protect\citeauthoryear{Erb et al.}{2006}]{Erb2006} Erb D.~K., Shapley A.~E., Pettini M., Steidel C.~C., Reddy N.~A., Adelberger K.~L., 2006, ApJ, 644, 813  
\bibitem[\protect\citeauthoryear{Feldmann}{2015}]{Feldmann2015} Feldmann R., 2015, MNRAS, 449, 3274  
\bibitem[\protect\citeauthoryear{Foreman-Mackey et al.}{2013}]{Foreman-Mackey2013} Foreman-Mackey D., Hogg D.~W., Lang D., Goodman J., 2013, PASP, 125, 306  
\bibitem[\protect\citeauthoryear{F{\"o}rster Schreiber \& Wuyts}{2020}]{ForsterSchreiber2020} F{\"o}rster Schreiber N.~M., Wuyts S., 2020, ARA\&A, 58, 661 
\bibitem[\protect\citeauthoryear{Fujimoto et al.}{2017}]{Fujimoto2017} Fujimoto S., Ouchi M., Shibuya T., Nagai H., 2017, ApJ, 850, 83  
\bibitem[\protect\citeauthoryear{Galliano, Galametz, \& Jones}{2018}]{Galliano2018} Galliano F., Galametz M., Jones A.~P., 2018, ARA\&A, 56, 673  
\bibitem[\protect\citeauthoryear{Garn \& Best}{2010}]{Garn2010} Garn T., Best P.~N., 2010, MNRAS, 409, 421  
\bibitem[\protect\citeauthoryear{Giannetti et al.}{2017}]{Giannetti2017} Giannetti A., et al., 2017, A\&A, 606, L12  
\bibitem[\protect\citeauthoryear{Goodman \& Weare}{2010}]{Goodman2010} Goodman J., Weare J., 2010, CAMCS, 5, 65  
\bibitem[\protect\citeauthoryear{Guo, Zheng, \& Fu}{2013}]{Guo2013} Guo K., Zheng X.~Z., Fu H., 2013, ApJ, 778, 23  
\bibitem[\protect\citeauthoryear{G{\'o}mez-Guijarro et al.}{2022}]{Gomez-Guijarro2022} G{\'o}mez-Guijarro C., et al., 2022, A\&A, 658, A43  
\bibitem[\protect\citeauthoryear{G{\'o}mez-Guijarro et al.}{2023}]{Gomez-Guijarro2023} G{\'o}mez-Guijarro C., Magnelli B., Elbaz D., Wuyts S., Daddi E., Le Bail A., Giavalisco M., et al., 2023, A\&A, 677, A34 

\bibitem[\protect\citeauthoryear{Hao et al.}{2011}]{Hao2011} Hao C.-N., Kennicutt R.~C., Johnson B.~D., Calzetti D., Dale D.~A., Moustakas J., 2011, ApJ, 741, 124  
\bibitem[\protect\citeauthoryear{Issa, MacLaren, \& Wolfendale}{1990}]{Issa1990} Issa M.~R., MacLaren I., Wolfendale A.~W., 1990, A\&A, 236, 237  
\bibitem[\protect\citeauthoryear{James et al.}{2002}]{James2002} James A., Dunne L., Eales S., Edmunds M.~G., 2002, MNRAS, 335, 753  
\bibitem[\protect\citeauthoryear{Johnson et al.}{2007}]{Johnson2007} Johnson B.~D., et al., 2007, ApJS, 173, 392  
\bibitem[\protect\citeauthoryear{Kaviraj}{2014}]{Kaviraj2014} Kaviraj S., 2014, MNRAS, 437, L41 
\bibitem[\protect\citeauthoryear{Katsianis et al.}{2020}]{Katsianis2020} Katsianis A., et al., 2020, MNRAS, 492, 5592  
\bibitem[\protect\citeauthoryear{Kennicutt \& Evans}{2012}]{Kennicutt2012} Kennicutt R.~C., Evans N.~J., 2012, ARA\&A, 50, 531  
\bibitem[\protect\citeauthoryear{Kennicutt}{1998}]{Kennicutt1998} Kennicutt R.~C., 1998, ApJ, 498, 541  
\bibitem[\protect\citeauthoryear{Kennicutt \& De Los Reyes}{2021}]{Kennicutt2021} Kennicutt R.~C., De Los Reyes M.~A.~C., 2021, ApJ, 908, 61  
\bibitem[\protect\citeauthoryear{Kong et al.}{2004}]{Kong2004} Kong X., Charlot S., Brinchmann J., Fall S.~M., 2004, MNRAS, 349, 769  
\bibitem[\protect\citeauthoryear{Kouroumpatzakis et al.}{2023}]{Kouroumpatzakis2023} Kouroumpatzakis K., Zezas A., Kyritsis E., Salim S., Svoboda J., 2023, A\&A, 673, A16  
\bibitem[\protect\citeauthoryear{Koyama et al.}{2019}]{Koyama2019} Koyama Y., Shimakawa R., Yamamura I., Kodama T., Hayashi M., 2019, PASJ, 71, 8  
\bibitem[\protect\citeauthoryear{Kreckel et al.}{2013}]{Kreckel2013} Kreckel K., et al., 2013, ApJ, 771, 62  
\bibitem[\protect\citeauthoryear{Kruijssen et al.}{2019}]{Kruijssen2019} Kruijssen J.~M.~D., Schruba A., Chevance M., Longmore S.~N., Hygate A.~P.~S., Haydon D.~T., McLeod A.~F., et al., 2019, Natur, 569, 519 
\bibitem[\protect\citeauthoryear{Krumholz \& McKee}{2005}]{Krumholz2005} Krumholz M.~R., McKee C.~F., 2005, ApJ, 630, 250  
\bibitem[\protect\citeauthoryear{Lada \& Dame}{2020}]{Lada2020} Lada C.~J., Dame T.~M., 2020, ApJ, 898, 3  
\bibitem[\protect\citeauthoryear{Lara-Lopez et al.}{2013}]{Lara-Lopez2013} Lara-Lopez M.~A., et al., 2013, MNRAS, 433, L35  
\bibitem[\protect\citeauthoryear{Larson}{1981}]{Larson1981} Larson R.~B., 1981, MNRAS, 194, 809  
\bibitem[\protect\citeauthoryear{Leja et al.}{2022}]{Leja2022} Leja J., et al., 2022, ApJ, 936, 165  
\bibitem[\protect\citeauthoryear{Leroy et al.}{2011}]{Leroy2011} Leroy A.~K., et al., 2011, ApJ, 737, 12  
\bibitem[\protect\citeauthoryear{Li et al.}{2019}]{Li2019} Li H., Wuyts S., Lei H., Lin L., Lam M.~I., Boquien M., Andrews B.~H., Schneider D.~P., 2019, ApJ, 872, 63  
\bibitem[\protect\citeauthoryear{Lisenfeld \& Ferrara}{1998}]{Lisenfeld1998} Lisenfeld U., Ferrara A., 1998, ApJ, 496, 145  
\bibitem[\protect\citeauthoryear{Liu et al.}{2013}]{Liu2013} Liu G., et al., 2013, ApJL, 778, L41  
\bibitem[\protect\citeauthoryear{Lofthouse et al.}{2017}]{Lofthouse2017} Lofthouse E.~K., Kaviraj S., Conselice C.~J., Mortlock A., Hartley W., 2017, MNRAS, 465, 2895 
\bibitem[\protect\citeauthoryear{Lombardi, Alves, \& Lada}{2010}]{Lombardi2010} Lombardi M., Alves J., Lada C.~J., 2010, A\&A, 519, L7  
\bibitem[\protect\citeauthoryear{Lorenz et al.}{2023}]{Lorenz2023} Lorenz B., et al., 2023, ApJ, 951, 29  
\bibitem[\protect\citeauthoryear{Lu et al.}{2023}]{Lu2023} Lu J., Shen S., Yuan F.-T., Zeng Q., 2023, ApJL, 946, L7  
\bibitem[\protect\citeauthoryear{Lu et al.}{2022}]{Lu2022} Lu J., Shen S., Yuan F.-T., Shao Z., Hou J., Zheng X., 2022, ApJ, 938, 139  
\bibitem[\protect\citeauthoryear{Maiolino \& Mannucci}{2019}]{Maiolino2019} Maiolino R., Mannucci F., 2019, A\&ARv, 27, 3  
\bibitem[\protect\citeauthoryear{Mannucci et al.}{2010}]{Mannucci2010} Mannucci F., Cresci G., Maiolino R., Marconi A., Gnerucci A., 2010, MNRAS, 408, 2115  
\bibitem[\protect\citeauthoryear{Martin et al.}{2005}]{Martin2005} Martin D.~C., et al., 2005, ApJL, 619, L59  
\bibitem[\protect\citeauthoryear{McKinnon et al.}{2018}]{McKinnon2018} McKinnon R., Vogelsberger M., Torrey P., Marinacci F., Kannan R., 2018, MNRAS, 478, 2851  
\bibitem[\protect\citeauthoryear{McLure et al.}{2018}]{McLure2018} McLure R.~J., et al., 2018, MNRAS, 476, 3991  
\bibitem[\protect\citeauthoryear{Misiriotis et al.}{2000}]{Misiriotis2000} Misiriotis A., Kylafis N.~D., Papamastorakis J., Xilouris E.~M., 2000, A\&A, 353, 117 
\bibitem[\protect\citeauthoryear{Mosenkov et al.}{2019}]{Mosenkov2019} Mosenkov A.~V., et al., 2019, A\&A, 622, A132  
\bibitem[\protect\citeauthoryear{Mowla et al.}{2019}]{Mowla2019} Mowla L.~A., et al., 2019, ApJ, 880, 57  
\bibitem[\protect\citeauthoryear{Naddaf \& Czerny}{2022}]{Naddaf2022} Naddaf M.~H., Czerny B., 2022, A\&A, 663, A77  
\bibitem[\protect\citeauthoryear{Nakajima et al.}{2023}]{Nakajima2023} Nakajima K., Ouchi M., Isobe Y., Harikane Y., Zhang Y., Ono Y., Umeda H., et al., 2023, ApJS, 269, 33 
\bibitem[\protect\citeauthoryear{Narayanan et al.}{2012}]{Narayanan2012} Narayanan D., Krumholz M.~R., Ostriker E.~C., Hernquist L., 2012, MNRAS, 421, 3127  
\bibitem[\protect\citeauthoryear{Nedkova et al.}{2021}]{Nedkova2021} Nedkova K.~V., et al., 2021, MNRAS, 506, 928  
\bibitem[\protect\citeauthoryear{Nelson et al.}{2013}]{Nelson2013} Nelson E.~J., et al., 2013, ApJL, 763, L16  
\bibitem[\protect\citeauthoryear{Nersesian et al.}{2019}]{Nersesian2019} Nersesian A., et al., 2019, A\&A, 624, A80  
\bibitem[\protect\citeauthoryear{Noeske et al.}{2007}]{Noeske2007} Noeske K.~G., et al., 2007, ApJL, 660, L43  
\bibitem[\protect\citeauthoryear{Oey et al.}{2003}]{Oey2003} Oey M.~S., Parker J.~S., Mikles V.~J., Zhang X., 2003, AJ, 126, 2317  
\bibitem[\protect\citeauthoryear{Ono et al.}{2023}]{Ono2023} Ono Y., et al., 2023, ApJ, 951, 72  
\bibitem[\protect\citeauthoryear{Pan et al.}{2021}]{Pan2021} Pan Z., Wang J., Zheng X., Kong X., 2021, ApJ, 922, 235  
\bibitem[\protect\citeauthoryear{Popesso et al.}{2023}]{Popesso2023} Popesso P., et al., 2023, MNRAS, 519, 1526  
\bibitem[\protect\citeauthoryear{Price et al.}{2014}]{Price2014} Price S.~H., et al., 2014, ApJ, 788, 86  
\bibitem[\protect\citeauthoryear{Qin et al.}{2019a}]{Qin2019a} Qin J., Zheng X.~Z., Wuyts S., Pan Z., Ren J., 2019, MNRAS, 485, 5733  
\bibitem[\protect\citeauthoryear{Qin et al.}{2019b}]{Qin2019b} Qin J., Zheng X.~Z., Wuyts S., Pan Z., Ren J., 2019, ApJ, 886, 28  
\bibitem[\protect\citeauthoryear{Qin et al.}{2022}]{Qin2022} Qin J., Zheng X.~Z., Fang M., Pan Z., Wuyts S., Shi Y., Peng Y., et al., 2022, MNRAS, 511, 765 
\bibitem[\protect\citeauthoryear{Rieke et al.}{2009}]{Rieke2009} Rieke G.~H., Alonso-Herrero A., Weiner B.~J., P{\'e}rez-Gonz{\'a}lez P.~G., Blaylock M., Donley J.~L., Marcillac D., 2009, ApJ, 692, 556  
\bibitem[\protect\citeauthoryear{Rosolowsky et al.}{2021}]{Rosolowsky2021} Rosolowsky E., et al., 2021, MNRAS, 502, 1218  
\bibitem[\protect\citeauthoryear{R{\'e}my-Ruyer et al.}{2014}]{Remy-Ruyer2014} R{\'e}my-Ruyer A., et al., 2014, A\&A, 563, A31  
\bibitem[\protect\citeauthoryear{Salim \& Narayanan}{2020}]{Salim2020} Salim S., Narayanan D., 2020, ARA\&A, 58, 529  
\bibitem[\protect\citeauthoryear{Sanders et al.}{2023}]{Sanders2023} Sanders R.~L., et al., 2023, ApJ, 942, 24  
\bibitem[\protect\citeauthoryear{Sanders et al.}{2021}]{Sanders2021} Sanders R.~L., et al., 2021, ApJ, 914, 19  
\bibitem[\protect\citeauthoryear{Sandstrom et al.}{2013}]{Sandstrom2013} Sandstrom K.~M., et al., 2013, ApJ, 777, 5  
\bibitem[\protect\citeauthoryear{Santoro et al.}{2022}]{Santoro2022} Santoro F., et al., 2022, A\&A, 658, A188  
\bibitem[\protect\citeauthoryear{Schreiber et al.}{2015}]{Schreiber2015} Schreiber C., et al., 2015, A\&A, 575, A74  
\bibitem[\protect\citeauthoryear{Schreiber et al.}{2018}]{Schreiber2018} Schreiber C., Elbaz D., Pannella M., Ciesla L., Wang T., Franco M., 2018, A\&A, 609, A30  
\bibitem[\protect\citeauthoryear{Shapley et al.}{2022}]{Shapley2022} Shapley A.~E., et al., 2022, ApJ, 926, 145  
\bibitem[\protect\citeauthoryear{Shapley et al.}{2023}]{Shapley2023} Shapley A.~E., Sanders R.~L., Reddy N.~A., Topping M.~W., Brammer G.~B., 2023, ApJ, 954, 157 
\bibitem[\protect\citeauthoryear{Shivaei et al.}{2020}]{Shivaei2020} Shivaei I., et al., 2020, ApJ, 899, 117  
\bibitem[\protect\citeauthoryear{Simard et al.}{2011}]{Simard2011} Simard L., Mendel J.~T., Patton D.~R., Ellison S.~L., McConnachie A.~W., 2011, ApJS, 196, 11  
\bibitem[\protect\citeauthoryear{Smith et al.}{2016}]{Smith2016} Smith M.~W.~L., et al., 2016, MNRAS, 462, 331  
\bibitem[\protect\citeauthoryear{Smith et al.}{2015}]{Smith2015} Smith R., Flynn C., Candlish G.~N., Fellhauer M., Gibson B.~K., 2015, MNRAS, 448, 2934 
\bibitem[\protect\citeauthoryear{Speagle et al.}{2014}]{Speagle2014} Speagle J.~S., Steinhardt C.~L., Capak P.~L., Silverman J.~D., 2014, ApJS, 214, 15  
\bibitem[\protect\citeauthoryear{Suess et al.}{2019}]{Suess2019} Suess K.~A., Kriek M., Price S.~H., Barro G., 2019, ApJ, 877, 103  
\bibitem[\protect\citeauthoryear{Tadaki et al.}{2020}]{Tadaki2020} Tadaki K.-. ichi ., et al., 2020, ApJ, 901, 74  
\bibitem[\protect\citeauthoryear{Thompson et al.}{2015}]{Thompson2015} Thompson T.~A., Fabian A.~C., Quataert E., Murray N., 2015, MNRAS, 449, 147  
\bibitem[\protect\citeauthoryear{Thorne et al.}{2021}]{Thorne2021} Thorne J.~E., et al., 2021, MNRAS, 505, 540  
\bibitem[\protect\citeauthoryear{Tremonti et al.}{2004}]{Tremonti2004} Tremonti C.~A., et al., 2004, ApJ, 613, 898  
\bibitem[\protect\citeauthoryear{Tuffs et al.}{2004}]{Tuffs2004} Tuffs R.~J., Popescu C.~C., V{\"o}lk H.~J., Kylafis N.~D., Dopita M.~A., 2004, A\&A, 419, 821  
\bibitem[\protect\citeauthoryear{van der Giessen et al.}{2022}]{vanderGiessen2022} van der Giessen S.~A., Leslie S.~K., Groves B., Hodge J.~A., Popescu C.~C., Sargent M.~T., Schinnerer E., Tuffs R.~J., 2022, A\&A, 662, A26  
\bibitem[\protect\citeauthoryear{van der Wel et al.}{2014a}]{van der Wel2014a} van der Wel A., et al., 2014, ApJL, 792, L6  
\bibitem[\protect\citeauthoryear{van der Wel et al.}{2014b}]{van der Wel2014b} van der Wel A., et al., 2014, ApJ, 788, 28  
\bibitem[\protect\citeauthoryear{van der Wel et al.}{2024}]{van der Wel2023} van der Wel A., Martorano M., H{\"a}u{\ss}ler B., Nedkova K.~V., Miller T.~B., Brammer G.~B., van de Ven G., et al., 2024, ApJ, 960, 53 
\bibitem[\protect\citeauthoryear{Whitaker et al.}{2012}]{Whitaker2012} Whitaker K.~E., van Dokkum P.~G., Brammer G., Franx M., 2012, ApJL, 754, L29  
\bibitem[\protect\citeauthoryear{Whitaker et al.}{2014}]{Whitaker2014} Whitaker K.~E., et al., 2014, ApJ, 795, 104  
\bibitem[\protect\citeauthoryear{Whitaker et al.}{2017}]{Whitaker2017} Whitaker K.~E., Pope A., Cybulski R., Casey C.~M., Popping G., Yun M.~S., 2017, ApJ, 850, 208  
\bibitem[\protect\citeauthoryear{Wild et al.}{2011}]{Wild2011} Wild V., Charlot S., Brinchmann J., Heckman T., Vince O., Pacifici C., Chevallard J., 2011, MNRAS, 417, 1760  
\bibitem[\protect\citeauthoryear{Wuyts et al.}{2011}]{Wuyts2011} Wuyts S., et al., 2011, ApJ, 738, 106  
\bibitem[\protect\citeauthoryear{Xiao et al.}{2012}]{Xiao2012} Xiao T., Wang T., Wang H., Zhou H., Lu H., Dong X., 2012, MNRAS, 421, 486  
\bibitem[\protect\citeauthoryear{Zahid et al.}{2012}]{Zahid2012} Zahid H.~J., Dima G.~I., Kewley L.~J., Erb D.~K., Dav{\'e} R., 2012, ApJ, 757, 54  
\bibitem[\protect\citeauthoryear{Zahid et al.}{2013}]{Zahid2013} Zahid H.~J., Yates R.~M., Kewley L.~J., Kudritzki R.~P., 2013, ApJ, 763, 92  
\bibitem[\protect\citeauthoryear{Zahid et al.}{2014}]{Zahid2014} Zahid H.~J., et al., 2014, ApJ, 792, 75  
\bibitem[\protect\citeauthoryear{Zahid et al.}{2017}]{Zahid2017} Zahid H.~J., Kudritzki R.-P., Conroy C., Andrews B., Ho I.-T., 2017, ApJ, 847, 18  
\bibitem[\protect\citeauthoryear{Zhang et al.}{2023}]{Zhang2023} Zhang J., et al., 2023, MNRAS, 524, 4128  
\bibitem[\protect\citeauthoryear{Zheng et al.}{2009}]{Zheng2009} Zheng X.~Z., et al., 2009, ApJ, 707, 1566  
\bibitem[\protect\citeauthoryear{Zhukovska}{2014}]{Zhukovska2014} Zhukovska S., 2014, A\&A, 562, A76  

\end{thebibliography}
\end{document}